\documentclass[11pt,a4paper]{article}
\usepackage{jcappub}

\usepackage{xcolor} 

\usepackage{arydshln} 

\newcommand{\msun}{\ensuremath{\mathrm{M}_\odot}}

\newcommand{\pnc}{\ensuremath{p_{N_c}}}
\newcommand{\pks}{\ensuremath{p_\text{K-S}}}

\DeclareMathOperator{\sech}{sech}

\title{Searching for a Galactic component in the IceCube track-like neutrino events}

\author[a]{Gregory S. Vance,}
\author[b,1]{Kimberly L. Emig,\note{Jansky Fellow of the National Radio Astronomy Observatory}}
\author[c]{Cecilia Lunardini,}
\author[a]{and Rogier A. Windhorst}

\affiliation[a]{School of Earth and Space Exploration, Arizona State University,\\Tempe, AZ 85287-1404, USA}
\affiliation[b]{National Radio Astronomy Observatory,\\520 Edgemont Road, Charlottesville, VA 22903, USA}
\affiliation[c]{Department of Physics, Arizona State University,\\Tempe, AZ 85287-1504, USA}

\emailAdd{Gregory.S.Vance@asu.edu}
\emailAdd{kemig@nrao.edu}
\emailAdd{Cecilia.Lunardini@asu.edu}
\emailAdd{windhors@asu.edu}

\abstract{
Searches for spatial associations between high-energy neutrinos observed at the IceCube Neutrino Observatory and known astronomical objects may hold the key to establishing the neutrinos' origins and the origins of hadronic cosmic rays.
While extragalactic sources like the blazar TXS~0506+056 merit significant attention, Galactic sources may also represent part of the puzzle.
Here, we explore whether open clusters and supernova remnants in the Milky Way contribute measurably to the IceCube track-like neutrino events above 200~TeV.
By searching for positional coincidences with catalogs of known astronomical objects, we can identify and investigate neutrino events whose origins are potentially Galactic.
We use Monte Carlo randomization together with models of the Galactic plane in order to determine whether these coincidences are more likely to be causal associations or random chance.
In all analyses presented, the number of coincidences detected was found to be consistent with the null hypothesis of chance coincidence.
Our results imply that the combined contribution of Galactic open clusters and supernova remnants to the track-like neutrino events detected at IceCube is well under 30\%.
This upper limit is compatible with the results presented in other Galactic neutrino studies.}

\begin{document}

\maketitle
\flushbottom

\section{Introduction} \label{intro}

The observed spectrum of cosmic rays extends to energies in excess of $10^{18}~\mathrm{eV}$ \cite{Beatty09}, raising fundamental questions about the nature of cosmic ray origins.
Especially above the spectral feature known as ``the knee'' at ${\sim}3 \times 10^{15}~\mathrm{eV}$ (3~PeV), it is still unknown where or how particles are accelerated to such high energies \cite[e.g.,][]{Blasi13, Gabici19}.
Cosmic rays at these energies are deflected by Galactic magnetic fields en route to Earth, scrambling their arrival directions and greatly complicating the task of source identification \cite{Halzen02}.

High-energy neutrino astronomy may hold the key to overcoming this problem, since cosmic accelerators are also expected to be astrophysical beam dumps. 
In an astrophysical beam dump, accelerated cosmic rays encounter ambient matter or photons in close proximity to the cosmic ray source, resulting in $pp$ or $p\gamma$ interactions which can produce neutral and charged pions ($\pi^0$, $\pi^\pm$) \cite{Gaisser90, Ahlers18}.
Subsequent pion decays yield gamma ray photons and neutrinos (as well as antineutrinos), which constitute secondary emission that can be used to trace the location of the accelerator \cite{Halzen02}.
Neutrinos are of particular interest, since they can reach Earth without deflection by magnetic fields and, in most cases, without flux attenuation due to intervening matter or background photons.

In 2013 and 2014, the IceCube Collaboration published the first evidence of high-energy extraterrestrial neutrinos \cite{IceCube13, Aartsen14a}.
A diffuse flux of neutrinos with energies of 0.01--2~PeV was detected, and its origins are still largely a mystery \cite[see, e.g.,][]{Aartsen17a, Aartsen17b, Aartsen19b, Albert20}.
With the exception of the blazar TXS~0506+056 \cite{IceCube18, Mirzoyan17, Ojha18} (possibly PKS~1502+106 \cite{IceCubeGCN19} and NGC~1068 \cite{Aartsen20a} as well), the majority of IceCube's neutrino events remain unattributed to any known astronomical sources.
Despite the prominence of the TXS~0506+056 result, an analysis by the IceCube Collaboration concluded that the combined diffuse neutrino flux contribution of all blazars listed in the 2nd Fermi-LAT AGN Catalog (2LAC) could be no more than 27\% \cite{Aartsen17c}, so the search continues for contributions from other types of sources.

In principle, the diffuse neutrino flux should have components of both Galactic and extragalactic origin.
Observations so far are consistent with isotropy, suggesting that any Galactic component is sub-dominant \cite{Aartsen17a}.
Still, it is plausible that ${\sim}10$--20\% of the flux originates within our Galaxy, with the overall neutrino spectrum having a softer Galactic component concentrated at energies $\lesssim 100~\mathrm{TeV}$ and a harder extragalactic component accounting for most of the highest-energy events \cite{Palladino16}.
It has been claimed that an excess of the IceCube neutrino flux above 100~TeV comes from the Galactic plane \cite{Neronov16}, but the IceCube Collaboration found that this excess flux is still compatible with the flux observed from other regions of the sky with a $p$-value of ${\sim}43\%$ \cite{Aartsen16}.
A more recent analysis by IceCube constrained the Galactic component to less than 14\% of the total flux above 1~TeV \cite{Aartsen17b}.
Other works, including those using the neutrino detector ANTARES, have found constraints that are either similar or even more restrictive ($< 9.5\%$ with 90\% confidence) \cite[e.g.,][]{Mandelartz15, Denton17, Albert17}.

Two main techniques have been used in attempts to associate high-energy neutrinos with Galactic or extragalactic sources: spatial association and neutrino emission modeling.
Studies searching for spatial associations often use statistical techniques to search for excesses in the number of neutrino events arriving from the directions of known objects or from pre-selected regions of the sky \cite[e.g.,][]{Emig15, Moharana15, Neronov16, Moharana16, Aartsen17c, Aartsen17b, Lucarelli19, Hooper19, Lunardini19, Kheirandish19, Moharana20}.
Since gamma ray emission due to $\pi^0$ decays is also expected from high-energy neutrino sources, gamma ray catalogs are often used to provide physically motivated lists of known objects for these types of analyses.
In the neutrino emission modeling approach, authors determine the expected neutrino flux from specific classes of sources (either individually or collectively) and compare their theoretical predictions with the empirical flux limits and spectral indices provided by neutrino observatories \cite[e.g.,][]{Yuan11, Ahlers14, Bednarek14, Anchordoqui14, Mandelartz15, Bykov15, Ahlers16, Bai16, Chakraborty16, Murase16, Halzen17, Gupta17, Bykov18, Palladino18, Sudoh18}.
Many such predictions are not yet testable and will require additional years of data or the increased sensitivity of future neutrino detectors such as KM3NeT \cite{AdrianMartinez16} or IceCube-Gen2 \cite{IceCubeGen214}.


In this work, we use a spatial association approach to search for a Galactic contribution among the track-like subset of neutrino events detected at IceCube \cite{Aartsen16}.
We restrict our Galactic search to two classes of energetic objects, both of which have seen consideration as potential high-energy neutrino sources: open clusters (OCs) \cite[e.g.,][]{Odrowski13, Bednarek14, Bykov15, Gupta17, Bykov18} and supernova remnants (SNRs) \cite[e.g.,][]{Yuan11, Mandelartz15, Ahlers16, Aartsen17b, Aartsen20c}.
The track-like IceCube neutrino events and our chosen astronomical catalogs of Galactic OCs and SNRs are discussed fully in Section~\ref{catalogs}.
The statistical analysis that we present is a variation of the well-known ``nearest neighbor'' method, which has seen previous use searching for coincidences with candidate neutrino sources \cite[e.g.,][]{Emig15, Moharana15, Lunardini19}.
In section~\ref{methods}, we discuss our method in detail and explain how its capabilities have been expanded for application to the Galactic plane.
Our results are presented and discussed in Section~\ref{results}, and we close with some brief conclusions in Section~\ref{conclusions}.
Our findings are broadly consistent with upper limits on Galactic sources as quoted by other studies \cite{Aartsen17b, Denton17, Albert17}.
The results act as a cross-check with previous SNR results by other authors, and also represent the first time that OCs have been considered in a spatial association neutrino search with a catalog of objects.

\section{Data} \label{catalogs}


\subsection{IceCube Track-like Neutrino Events} \label{nu_data}

The IceCube Neutrino Observatory is made up of of 5160 photomultiplier tubes frozen into $1~\mathrm{km}^3$ of Antarctic ice at the South Pole \cite{Aartsen17d}.
When a high-energy neutrino (or antineutrino) of any flavor interacts with an atomic nucleus in the ice, relativistic charged particles are produced, emitting Cherenkov light that IceCube detects \cite{Aartsen17e}.
Neutrino events in the detector have either a track-like signature or a cascade signature; the latter are produced by neutral-current interactions of all three neutrino flavors, as well as by charged-current interactions of electron and tau neutrinos \cite{Aartsen17e}.
Cascades yield particle showers of range $\lesssim 20~\mathrm{m}$ \cite{Aartsen14c} with a light signature that is nearly spherically symmetric, so the angular uncertainty of cascade event reconstructions is $\gtrsim 10^\circ$ \cite{Aartsen17e}.
In contrast, a charged-current interaction of a muon neutrino yields a high-energy muon which typically travels several kilometers through the ice \cite{Chirkin04}, producing a track-like signature with typical angular resolution of $\lesssim 1^\circ$ in reconstruction \cite{Aartsen17e}.

Our analysis uses 29 track-like neutrino events published by the IceCube Collaboration after six years of operation \cite{Aartsen16}.
All of these events have observed energies of 200~TeV or more, an imposed energy cut which results in an expectation of roughly twice as many astrophysical events as background atmospheric events.
Each neutrino event is published with angular errors on the reconstructed arrival direction corresponding to 50\% and 90\% confidence levels that are asymmetric in both right ascension and declination.
For the purposes of our statistical analysis, we create circular error regions centered on each event with the same area as the original error ellipse by using the geometric mean of all four angular errors at the 90\% confidence level (see Section~\ref{the_method}).
We use circularized errors because our statistical method is primarily sensitive to the total area of the error ellipse, and circular errors simplify the angular separation calculations required.
Note that we do not consider cascade events in this work due to their large angular uncertainties---the smaller errors of the track-like events make them better-suited to point source searches using positional coincidence, since chance associations become less likely when the collection of all error ellipses covers a smaller solid angle on the sky.

\begin{figure}
\centering
\includegraphics[width=5.5in]{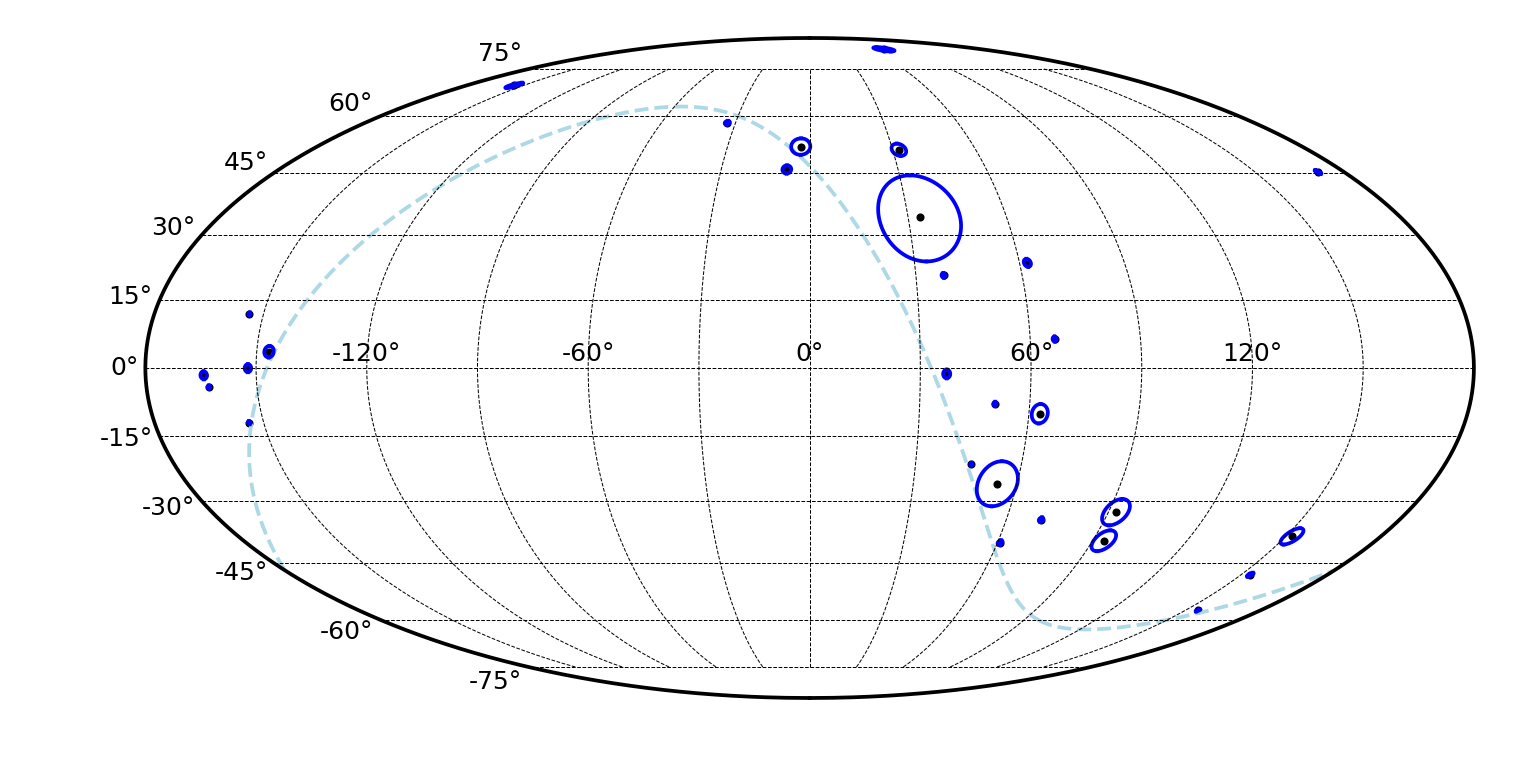}
\caption{Sky map in Galactic coordinates showing the distribution of arrival directions (black dots) and angular errors (blue circles) for the 29 track-like neutrinos \cite{Aartsen16}.
The displayed errors have been circularized as described in the text.
The dashed light blue line shows the celestial equator.}
\label{track_skyplot}
\end{figure}

The reconstructed arrival directions of the 29 track-like neutrinos are concentrated in the lower latitudes of the Northern Hemisphere (declinations $-5^\circ < \delta < +50^\circ$, see Figure~\ref{track_skyplot}).
Most datasets published by the IceCube Collaboration require that the vertex of neutrino interaction occur within the instrumented detector volume, but this set of track-like events allows for interactions both inside and outside the detector.
This increases the effective area of the detector due to the long range of the muons produced by muon neutrino interactions, but makes it necessary to restrict the field of view to the Northern Hemisphere (below IceCube's horizon) where the Earth efficiently filters out the background noise of atmospheric muons.

\subsection{Candidate Sources: Open Clusters} \label{oc_data}

OCs are one class of Galactic source which has been suspected of playing host to high-energy cosmic accelerators.
Many such OCs represent compact collections of dense interstellar matter grouped together with energetic objects such as massive OB stars, compact massive binaries, pulsars (often with associated pulsar wind nebulae), and SNRs \cite{Bednarek14}. 
Several of these types of objects have been investigated as suspected cosmic ray accelerators by themselves.
For instance, Galactic SNRs are likely the principal source of cosmic rays at energies below the knee (see Section~\ref{snr_data}). 
The Eta Carinae binary system in Trumpler~16 has been detected in gamma-ray observations up to ${\sim}100\ \mathrm{GeV}$ and has been studied as a potential accelerator \cite{Tavani09, Abdo09, Farnier11, Bednarek11, Gupta17}.
Massive X-ray binaries (microquasars) including Cyg~X-3 and LS~5039 have been proposed as Galactic neutrino emitters \cite{Bednarek05, Anchordoqui14}.
A few notable star clusters have been found to produce TeV gamma-ray emission, including Cygnus~OB2 \cite{Aharonian02} and the larger Cygnus~X star-forming region \cite{YoastHull17, Abeysekara21}, Westerlund~2 \cite{Aharonian07}, Westerlund~1 \cite{Abramowski12}, and Cl*~1806-20 \cite{HESS18}.
Such emission could be indicative of $\pi^0$ decays associated with hadronic neutrino production.
Several theoretical studies of star clusters have examined neutrino detection prospects, and the results suggest fluxes detectable by IceCube or by next-generation detector KM3NeT, especially if several acceleration mechanisms are simultaneously active within the same cluster \cite{Bednarek14, Bykov15, Bykov18}.

A wide variety of physical accelerator scenarios have been proposed in order to explain the observed TeV gamma-ray emission from select OCs.
These scenarios frequently involve powerful winds from massive stars, supernova shocks from the deaths of those stars, and/or compact objects from the collapsed cores of exploded stars interacting with dense regions of ambient gas and dust within the star cluster \cite[e.g.,][and references therein]{Bednarek14}.
In massive binary systems, particle acceleration may occur in the region where the two stars' stellar winds collide \cite{Bednarek11, Bednarek14, Gupta17}.
Galactic microquasars are powered by accretion disks whose associated jets could transfer considerable kinetic energy to relativistic nuclei \cite{Bednarek05, Anchordoqui14}.
The collision of a supernova shock with the cumulative stellar winds of a compact cluster of young stars might theoretically be capable of accelerating protons to energies in excess of 1~PeV \cite{Bykov15, Bykov18}.
At this time, it is unclear which of these processes (if any) contribute most to the high-energy budget of a star cluster; indeed, the answer may vary from one cluster to the next \cite{Bednarek14}.

For our investigation of OCs, we employ two datasets: (1) an extensive, general catalog of Galactic OCs and (2) a more restricted list of massive, young OCs which may provide an optimal accelerator environment for producing neutrinos and gamma-ray emission.
Below, we briefly describe each list of candidate sources and our motivations for selecting those objects.

\begin{itemize}

\item \textit{New catalog of optically visible open clusters and candidates} \cite{Dias02}

OPENCLUST is a list of Galactic OCs which was first compiled in 2002 from a combination of older catalogs \cite{Lauberts82, Lynga87, Mermilliod95} and several isolated papers published more recently \cite[for a full list, see][]{Dias02}.
Many objects were visually checked by the authors using Digitized Sky Survey\footnote{\url{https://archive.stsci.edu/dss/}} (DSS) plates in order to verify their coordinates and existence.
The list includes fundamental parameters and kinematic data where available, and is designed to be an efficient tool for use in OC studies, since all of the included data is stored in a single machine-readable table.
The catalog is regularly updated online\footnote{HEASARC implementation: \url{https://heasarc.gsfc.nasa.gov/W3Browse/all/openclust.html}}---the version used in this work contains entries for a total of 2167 OCs and OC candidates, at least 89\% of which represent robustly identified OCs.

This extensive collection of objects is useful for testing a broad hypothesis of potential causal association that is more model-independent: if there is a strong causal connection between IceCube neutrinos and Galactic OCs---regardless of the underlying physical reasons---this catalog could reveal the existence of that connection.
In addition, OPENCLUST lists the apparent angular size of each OC, which is important for our statistical analysis.
OPENCLUST will also provide the framework for constructing a probabilistic model of the spatial distribution of OCs in the Milky Way (see Section~\ref{galactic_null}); its sheer size makes it an ideal data set for this task.

\item \textit{``Accelerator-dominated'' open clusters} \cite{Odrowski13}

This data set consists of 36 ``accelerator-dominated'' open clusters (ADOCs) compiled in 2013 by the IceCube collaboration.
The ADOCs are selected from among the 650 Galactic OCs documented in the Catalog of Open Cluster Data (COCD) \cite{Kharchenko05a} and its First Extension \cite{Kharchenko05b}.
The authors limit their study to sources in the Northern Hemisphere, since that is where the sensitivity of IceCube is highest, and they also do not consider sources farther than 3--4~kpc from the Sun, with the goal of eliminating OCs with imprecise distance estimates or identifications.
The primary selection criterion used to identify OCs in an ``accelerator-dominated'' phase was their evolutionary status---to be included, an OC must have a turn-off mass above 9~\msun{}, corresponding to an age of less than ${\sim}40$~Myr.
OCs much older than this are assumed to be deficient in potential particle accelerators since massive stars will have died and their supernovae will have long since dispersed into the interstellar medium.
The final list includes a range of young OCs with total masses between ${\sim}400$~\msun{} and $3 \times 10^4$~\msun{}, which are largely concentrated in three cluster complexes associated with spiral arms of the Galaxy.

The ADOC list is a narrow set of sources assembled with the specific aim of using IceCube to test the hypothesis of young OCs as cosmic ray accelerators and high-energy neutrino emitters.
If cosmic ray acceleration within OCs contributes significantly to a Galactic neutrino flux component at IceCube, then these 36 OCs are the most likely candidates for detection.
Like OPENCLUST, the COCD which the ADOCs were selected from also provides apparent angular sizes of each OC for our analysis.
Finally, we note that Table~1 of \cite{Odrowski13} lists IC~1448 among the 36 selected ADOCs, but this object is a galaxy (better known as NGC~7308) and is not present in the COCD.
We will assume that the authors meant to refer to IC~1848 instead.

\end{itemize}

\subsection{Candidate Sources: Supernova Remnants} \label{snr_data}

As stated in Section~\ref{oc_data}, Galactic SNRs are likely to be the primary source of hadronic cosmic rays with energies at least up to ${\sim}100\ \mathrm{TeV}$, as indicated by a large yet circumstantial body of evidence \cite{Yuan11, Blasi13, Bykov15}.
Gamma-ray spectra observed from several SNRs including W44 \cite{Ackermann13}, IC~443 \cite{Ackermann13, Tavani10}, Cassiopeia~A \cite{Araya10}, Tycho's SNR \cite{Morlino12, Berezhko13}, and others \cite[e.g.,][]{Ohira11} have been found to favor hadronic emission models (i.e., $\pi^0$ decay) over leptonic explanations such as inverse Compton.
There are also SNRs such as RX~J1713.7$-$3946 \cite{Abdo11} which have gamma-ray spectra that are more compatible with leptonic models.
Therefore, the observation of a neutrino flux from SNRs with hadronic gamma-ray emission would offer another tool for differentiating between the two scenarios.
The presence of accelerated particles in SNR shocks has been inferred in many cases from X-ray evidence of magnetic field amplification \cite{Volk05, Vink12, Bykov12, Bykov13, Schure12} and observations of unusual $\mathrm{H}\alpha$ line widths \cite{Blasi13, Heng10}.


The common theoretical framework for cosmic ray acceleration in SNRs is diffusive shock acceleration (DSA) at the supernova shock \cite[for a review, see][]{Bell14}.
The general predictions of DSA are a power law spectrum $E^{-\Gamma}$ of cosmic ray energies with spectral index $\Gamma \approx \text{2.0--2.4}$ up to a cutoff energy, which might extend as high as the knee in some cases \cite{Villante08}.
As cosmic ray energies approach PeV scales, the role of Galactic SNRs becomes less certain; the ability of DSA to reach energies as high as the knee has been repeatedly questioned \cite[e.g.,][]{Bell14, Bykov15}.
Hence, a number of more complex scenarios have been proposed.
It has been argued that only certain kinds of supernovae \cite{Ptuskin10, Bell14} or supernovae exploding in particular environments \cite{Bykov15, Binns05} provide the conditions necessary to reach knee energies.
Magnetic field amplification is likely to be one piece of the puzzle \cite{Schure12}, and the answer may also include plasma instabilities or alternative shock arrangements \cite[e.g.,][and references therein]{Bell14, Bykov15}.
The SNR G40.5$-$0.5 has been proposed as a potentially detectable point source in IceCube given sufficient exposure over a period of ${\sim}10\ \mathrm{yr}$ \cite{Mandelartz15}.
A few other Galactic SNRs, including Cassiopeia~A, IC~443, Vela Junior, W33, and W41, have also been proposed as neutrino emitters, but theoretical predictions do not expect these to be detectable as point sources in IceCube above the atmospheric backgrounds \cite{Yuan11, Mandelartz15}.

For our SNR study, we make use of another pair of catalogs: (1) a comprehensive list of every known Galactic SNR and (2) a limited subset of SNRs which have been detected at gamma-ray energies and are therefore more likely to play host to high-energy accelerators.
The remainder of this subsection describes both of these catalogs and justifies their use.

\begin{itemize}

\item \textit{Catalogue of Galactic supernova remnants by D.~A. Green} \cite{Green14}

The Green SNRs catalog is a compilation of all currently known Galactic SNRs, which has seen numerous published versions since 1984 \cite[for a list, see][]{Green14}.
Here, we make use of the updated June 2017 version \cite{Green17}, which is available online.\footnote{See \url{http://www.mrao.cam.ac.uk/surveys/snrs/}.}
The catalog lists 295 Galactic SNRs together with observational data on each object's position, angular size, morphology, radio emission spectral index, and derived flux density at a frequency of 1~GHz.
Most of these objects have been detected through their radio emission due to relativistic electrons, which is free of Galactic extinction.
More than 93\% of the catalog's SNRs have been sufficiently observed in the radio that their flux density at 1~GHz can be determined and included.
Due to Galactic absorption at other wavelengths, only ${\sim}40\%$ of the SNRs have been detected in X-rays and ${\sim}30\%$ in the optical.
The catalog's completeness is influenced by some selection effects, primarily those affecting radio wavelengths \cite[e.g.,][]{Green15}.
The catalog is complete only down to a surface brightness of ${\sim}10^{-20}\ \mathrm{W\ m^{-2}\ Hz^{-1}\ sr^{-1}}$ at 1~GHz, and a higher limit likely applies near the Galactic center, where the background Galactic radio emission is brightest \cite{Green14, Green15}.

Despite any limitations, the Green SNRs catalog is the largest available listing of Galactic SNRs.
Much like OPENCLUST, the sheer number of objects present makes this collection ideal for testing the broadest class of hypotheses that connects IceCube neutrinos to Galactic SNRs.
The angular size estimates in the catalog are relevant for our statistical analysis as presented in Section~\ref{methods}.
We will also use the distribution of these SNRs to generate our probabilistic model of the distribution of SNRs in the Milky Way.

\item \textit{First Fermi LAT supernova remnant catalog} \cite{Acero16}

The Fermi LAT SNRs are the result of a systematic effort to characterize the 1--100~GeV gamma ray emission within $3^\circ$ of all known radio SNRs from the Green SNR catalog \cite{Green14, Green17}.
A total of 30 candidate sources were found to have sufficient significance and spatial overlap with known radio SNRs to be classified by the authors as GeV counterparts of the associated radio SNRs.
An additional 14 candidates were designated as ``marginally classified'' due to lower degrees of significance and/or spatial overlap, and a further 245 flux upper limits were presented in other cases.
By scrambling the Galactic longitudes of known SNRs to create a mock catalog, the authors determined an upper limit of 22\% for the fraction of candidates that were falsely identified as SNRs.

In this work, we use both the classified and marginally classified candidates from the Fermi LAT SNRs catalog for a grand total of 44 objects.
The inclusion of the marginally classified candidates did not qualitatively change our results.
The catalog provides positional data and angular radius estimates for each object, both of which are used for our analysis as presented in Section~\ref{methods}.
Much like the ADOC list presented in Section~\ref{oc_data}, this catalog has the advantage that it carries greater physical motivation in terms of cosmic ray acceleration and neutrino emission.
Since gamma-ray emission can result from the same hadronic processes, these Galactic SNRs are more likely to be neutrino emitters than most of the Green SNRs.

\end{itemize}

\section{Methodology} \label{methods}

\subsection{Statistical Method of Coincidences} \label{the_method}

Our analysis technique is a variant of the well-established ``nearest neighbor'' method.
Such methods have a history of use in both astronomy \cite[e.g.,][]{deRuiter77, Windhorst84, Sutherland92} and astroparticle physics \cite[e.g.,][]{Virmani02, Moharana15, Emig15, Lunardini19}.
Here, we summarize the important concepts, focusing mainly on the newest additions.
We refer to our previous work \cite{Emig15, Lunardini19} for a more extended presentation.

We begin with a collection of $N$ neutrino events and $M$ candidate sources.
For each event $\nu_i$, where $i = 1, \ldots, N$, we define the unitless quantity $r_i$ as
\begin{equation}
r_i = \min\limits_{j} \left( \frac{S_{i,j} - a_j}{\sigma_{\nu,i}} \right) ,
\end{equation}
where $j = 1, \ldots, M$ runs over the set of candidate sources.
The quantity $S_{i,j}$ is the angular separation between $\nu_i$ and the center of the $j$th candidate source, $a_j$ is the angular radius of the $j$th candidate source, and $\sigma_{\nu,i}$ is the angular error associated with event $\nu_i$.
To simplify the calculation of $r_i$, the neutrino event error ellipses and the candidate sources are both treated as circular (which may cause individual values of $r_i$ to be underestimated or overestimated).
Where necessary, their shapes were circularized using the geometric mean of the available dimensions to yield a single angular radius for a circle with the same area as the original ellipse.\footnote{By approximating small circles and ellipses on the spherical sky with their Euclidean equivalents, we can show that $A_\mathrm{circle} = \pi r^2 = \pi (\sqrt{a b})^2 = \pi a b = A_\mathrm{ellipse}$.}
If the angular extent of a candidate source is unavailable, then we take the conservative approach of treating it as a point source with $a_j = 0$.

The quantity $r_i$ can be regarded as the normalized angular separation between the arrival direction of $\nu_i$ and the edge of the nearest candidate source.
From here onward, the index $i$ will be dropped for simplicity, and the notation $r$ will be used to indicate the normalized angular separation.
For any catalog of candidate source positions and angular sizes, we can empirically calculate the $r$ values, one for each of the $N$ neutrino events.
To test for a causal relationship between the candidate sources and the neutrinos, we compare the distribution of $r$ values with a null distribution, which is the expected distribution of $r$ values in a scenario where there exists no causal relationship and any apparent correlations are due to random chance.
The null distribution is obtained using a Monte Carlo procedure which generates $K = 10^6$ randomized realizations of the $N$ values for $r$.

Our primary strategy to compare the two distributions is to use the number of coincidences $N_c$ as a test statistic.
We define a \textit{coincidence} as the occurrence of a neutrino event having $r < 1$, so the number of coincidences is an integer such that $0 \leq N_c \leq N$.
Many of the $K$ Monte Carlo realizations of the null distribution will include coincidences that occur by random chance.
We define $N_\mathrm{better}$ to be the number of Monte Carlo realizations which happen to produce at least $N_c$ coincidences.
Then, $p = N_\mathrm{better} / K$ is our $p$-value, the approximate probability that a number of coincidences larger than or equal to the observed $N_c$ is realized in the null distribution.
If $N_c$ greatly exceeds the typical number of coincidences expected in the null hypothesis, then $p \ll 1$, which could be indicative of a causal relationship between the neutrinos and the candidate sources.

As an alternative approach, we also study the entire distribution of $r$ using a Kolmogorov-Smirnov (K-S) test.
Using a two-sample K-S test, we evaluate the hypothesis that the $N$ empirical $r$ values were drawn from the same underlying distribution as the set of $K N$ values drawn from the null distribution.
If the test's $p$-value is near zero, it suggests that the empirical $r$ values were drawn from an underlying distribution which differs from the null distribution in a statistically significant way.
Such a result could indicate a relationship between the chosen candidate sources and the neutrinos, but not necessarily a causal one.
While the first approach using the number of coincidences is a more direct way of discovering spatial correlations in the data, the K-S test approach is a more general check for consistency between the real data and the null distribution.
We note that the $p$-values resulting from both strategies are pre-trial values, since they do not take into account the total number of hypotheses that were tested.
Post-trial $p$-values would be larger, i.e., they would have lower significance.

\subsection{Galactic Null Distributions} \label{galactic_null}

The Monte Carlo procedure for generating null distribution realizations approaches the problem by randomizing the positions of the $M$ candidate sources while leaving the arrival directions of the $N$ neutrino events fixed.
Extragalactic sources can, in most cases, be assumed to have an isotropic distribution on the sky \cite{Emig15, Lunardini19}, which we imitate by choosing points uniformly distributed on the surface of a sphere.
Since the candidate sources used in this work are Galactic sources, they are \textit{not} distributed isotropically, but are instead heavily concentrated along the Galactic plane.
For each candidate source catalog that we use, 96\% or more of its objects are within $20^\circ$ of the Galactic plane, i.e., they have Galactic latitudes $b$ satisfying $-20^\circ < b < +20^\circ$.
In addition, the SNR catalogs show a greater concentration of objects towards the Galactic center at Galactic longitude $\ell = 0^\circ$ and a correspondingly lower concentration towards the Galactic anticenter at $\ell = \pm 180^\circ$.
Therefore, the Monte Carlo randomization must take all these physical realities into account.

Using the procedure known as transform sampling \cite[e.g., see Section~7.2 of][]{Press92}, random values are generated from any desired probability distribution.
To model the probability distribution function (PDF) of Galactic objects on the sky, we used Bayesian parameter estimation with the \verb|dynesty| software package \cite{Speagle20} to obtain median posterior parameter values for a set of model distributions.
Since \verb|dynesty| also provides estimates of the Bayesian evidence, we were able to compare the models to one another using odds ratios and select those which showed the best agreement with the available data.
We consider a factorized PDF having the form $F(b, \ell) = B(b) L(\ell)$, where $B(b)$ and $L(\ell)$ are assumed to be independent functions.
This is accurate to first order, since the latitude and longitude distributions of objects in the Galactic disk arise from two distinct physical mechanisms \cite{Binney97, Oort32}, even if a mild interdependence may exist on small scales.

For the latitude distribution $B(b)$, the best model (as determined via odds ratios) used a functional form inspired by the plane-perpendicular blue-light distribution of edge-on galaxy disks \cite{deGrijs97}:\footnote{Note that $\sech x$ is the hyperbolic secant of $x$, defined as $\sech x \equiv 1 / \cosh x = 2 / (e^x + e^{-x})$, where $\cosh$ is the hyperbolic cosine.}
\begin{equation}
B(b) = A_b \left[ f_1 \sech\left( \frac{b - b_0}{h_{b1}} \right) + f_2 \sech\left( \frac{b - b_0}{h_{b2}} \right) \right] \qquad f_1 + f_2 = 1 \qquad f_1, f_2 \geq 0 . \label{lat_dist}
\end{equation}
This model surpassed eight other models including a cosine, an exponential, a single hyperbolic secant, a hyperbolic secant squared, a Gaussian, and the full functional form proposed for edge-on galaxy disks \cite{deGrijs97}.
The four independent parameters are $b_0$, $h_{b1}$, $h_{b2}$, and $f_1 = 1 - f_2$.
The value $A_b$ is the normalization, chosen so that $B(b)$ integrates to 1 over the interval $-90^\circ \leq b \leq +90^\circ$.
The parameter $b_0$ represents the offset position of the central peak, about which the distribution is symmetric.
Such an offset from the Galactic plane is expected due to the vertical position of the Sun within the Galactic disk.
The parameter $h_{b1}$ is the (relatively narrow) angular width of the first $\sech$ distribution (representing the thin disk of the Galaxy), and $h_{b2}$ is the (relatively broad) angular width of the second $\sech$ distribution (representing the thick disk).
The parameters $f_1$ and $f_2$ are the fractional amplitudes of the two $\sech$ distributions, which together comprise a single independent parameter.

The longitude distribution $L(\ell)$ varied by catalog.
For some catalogs, the dependence on longitude was weak enough that the best model was a uniform distribution with the functional form
\begin{equation}
L(\ell) = \frac{1}{360^\circ} . \label{lon_dist_flat}
\end{equation}
This model has zero parameters, and the constant value $1/360^\circ$ is the necessary normalization so that $L(\ell)$ integrates to 1 over the interval $-180^\circ \leq \ell < +180^\circ$.
Other catalogs were better modeled by a standard exponential disk model \cite[e.g., see Section~2.2 of][]{Sparke07} of the form
\begin{equation}
L(\ell) = A_\ell \exp \left( \frac{-|\ell - \ell_0|}{h_\ell} \right) , \label{lon_dist_exp}
\end{equation}
where the periodic longitude angle is always chosen such that $-180^\circ \leq \ell - \ell_0 < +180^\circ$.
The two independent parameters are $\ell_0$ and $h_\ell$, with the value $A_\ell$ serving as the normalization.
The $\ell_0$ parameter represents any necessary offset between the central peak of the distribution and the Galactic center at $\ell = 0^\circ$, and $h_\ell$ is an angular scale height related to the radial scale height of the Galactic disk.
No additional longitude distribution models were considered.

\begin{figure}
\centering
\includegraphics[height=1.8in]{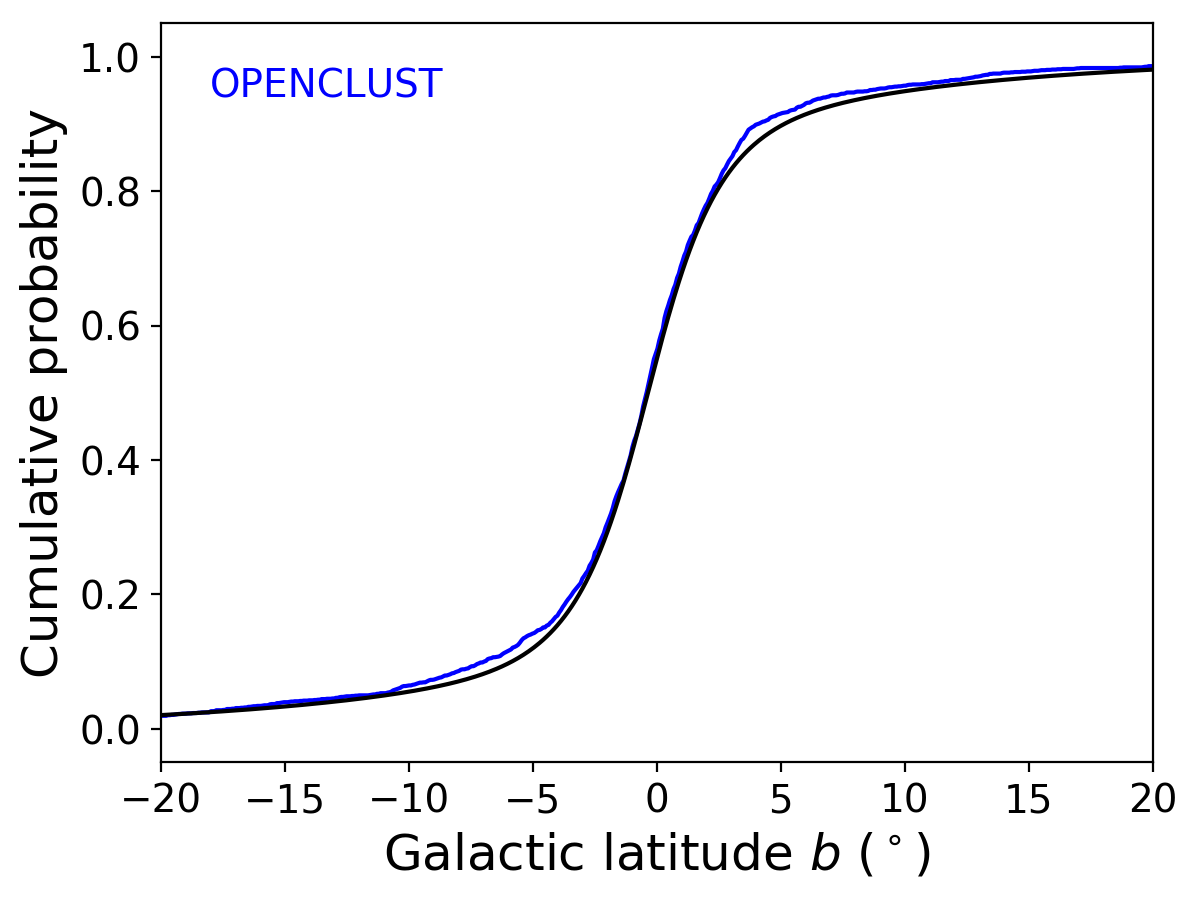}
\includegraphics[height=1.8in]{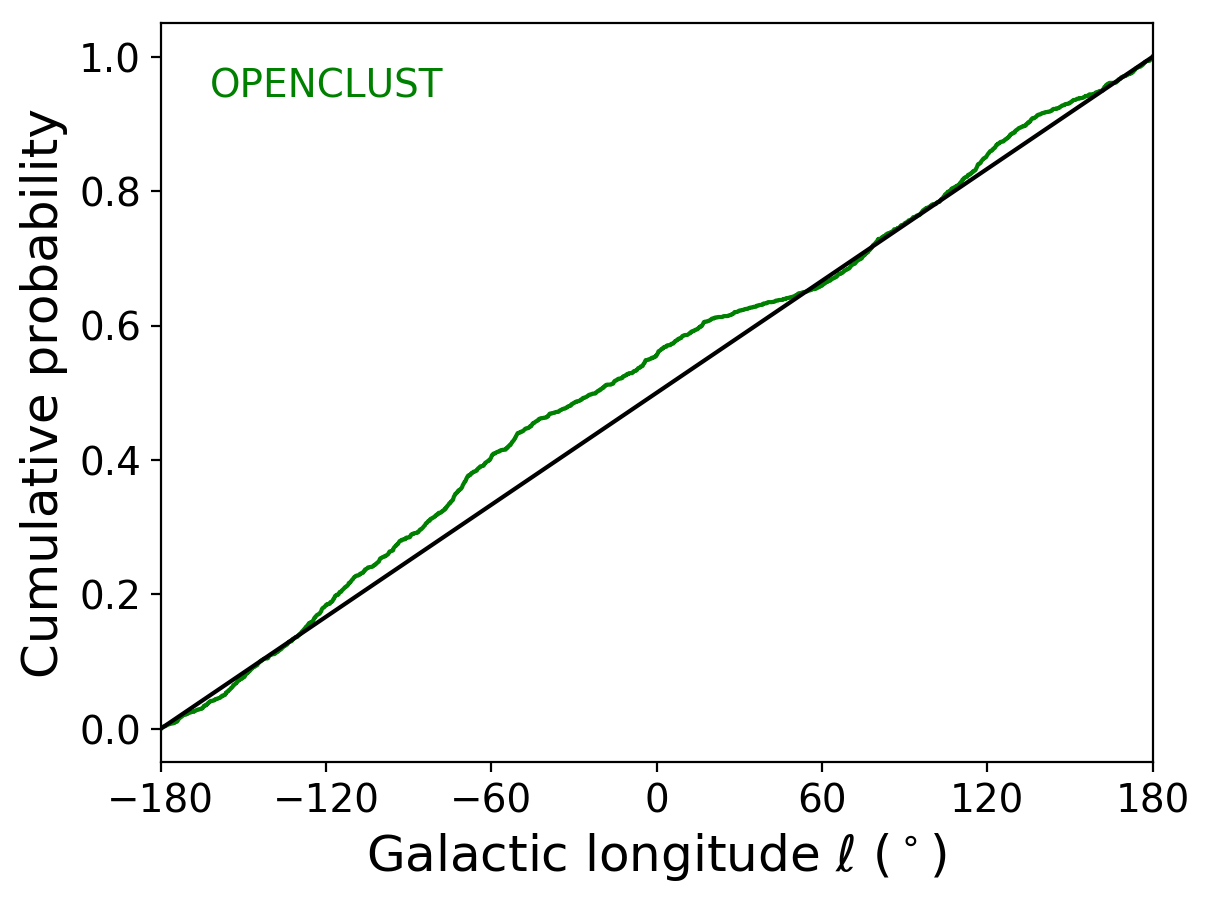}
\includegraphics[height=1.8in]{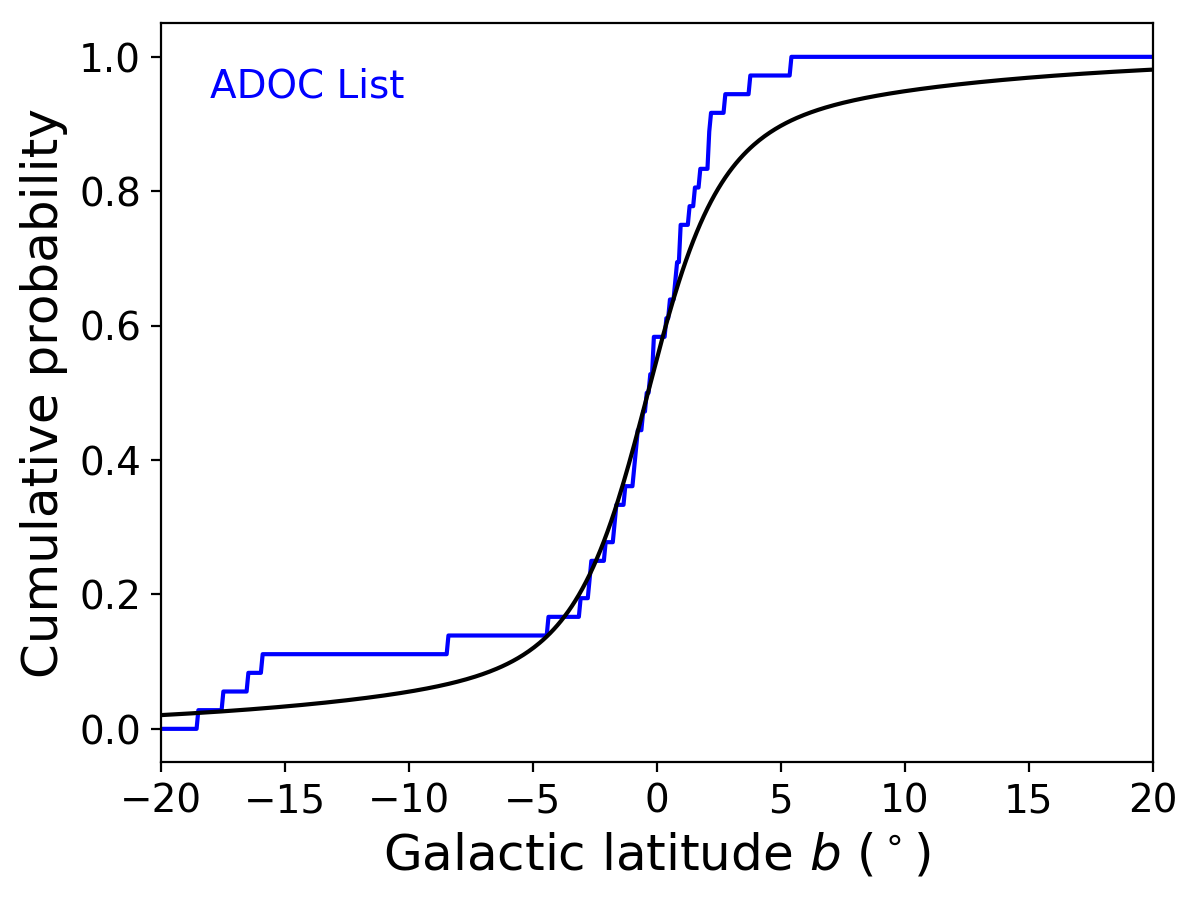}
\includegraphics[height=1.8in]{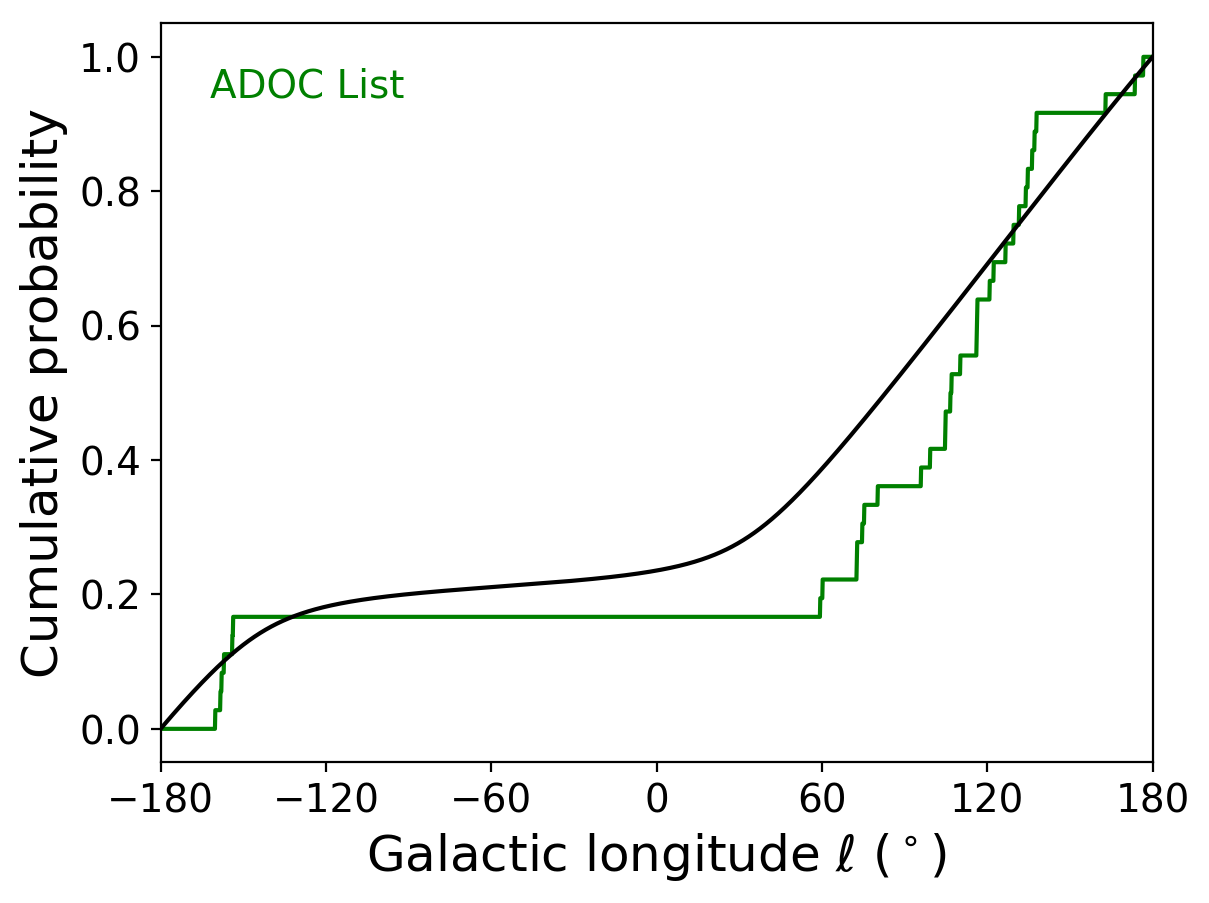}
\includegraphics[height=1.8in]{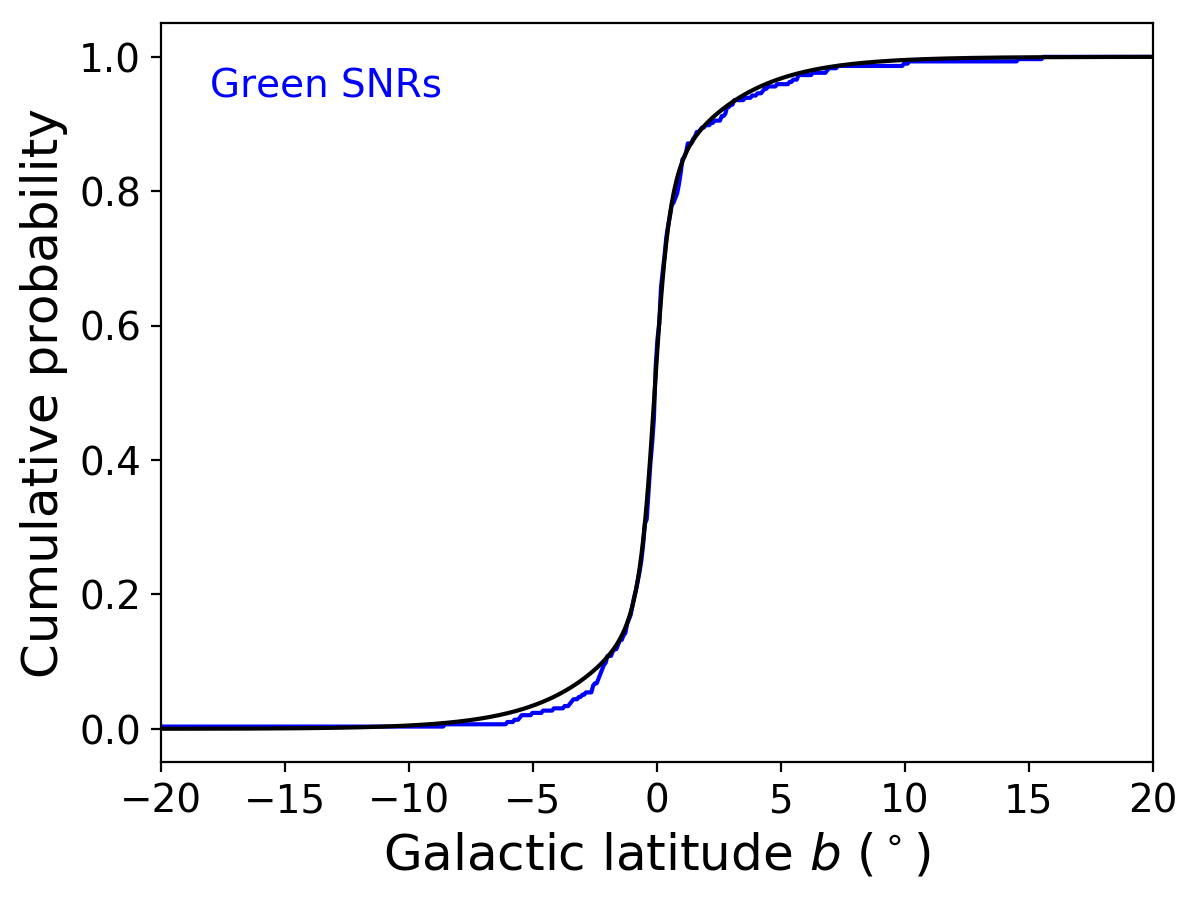}
\includegraphics[height=1.8in]{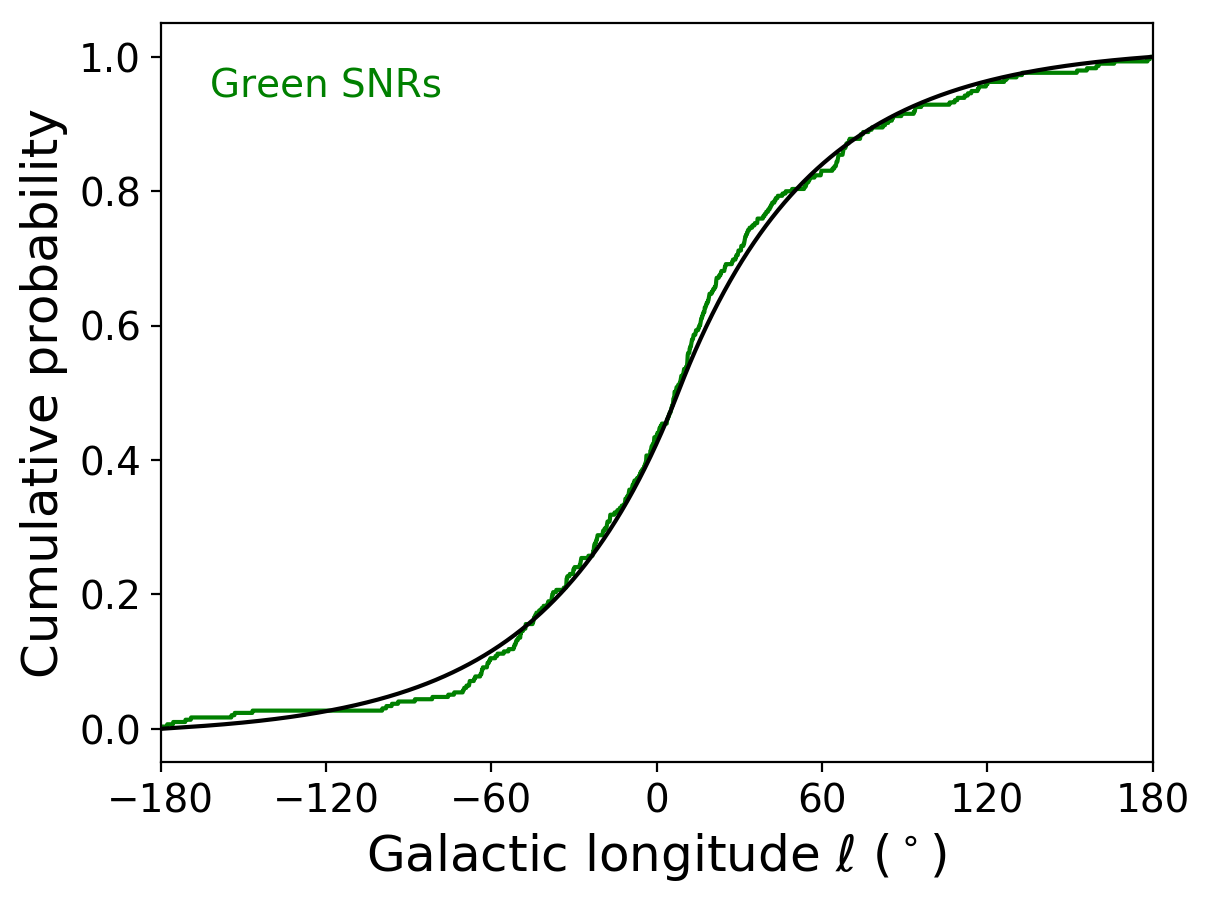}
\includegraphics[height=1.8in]{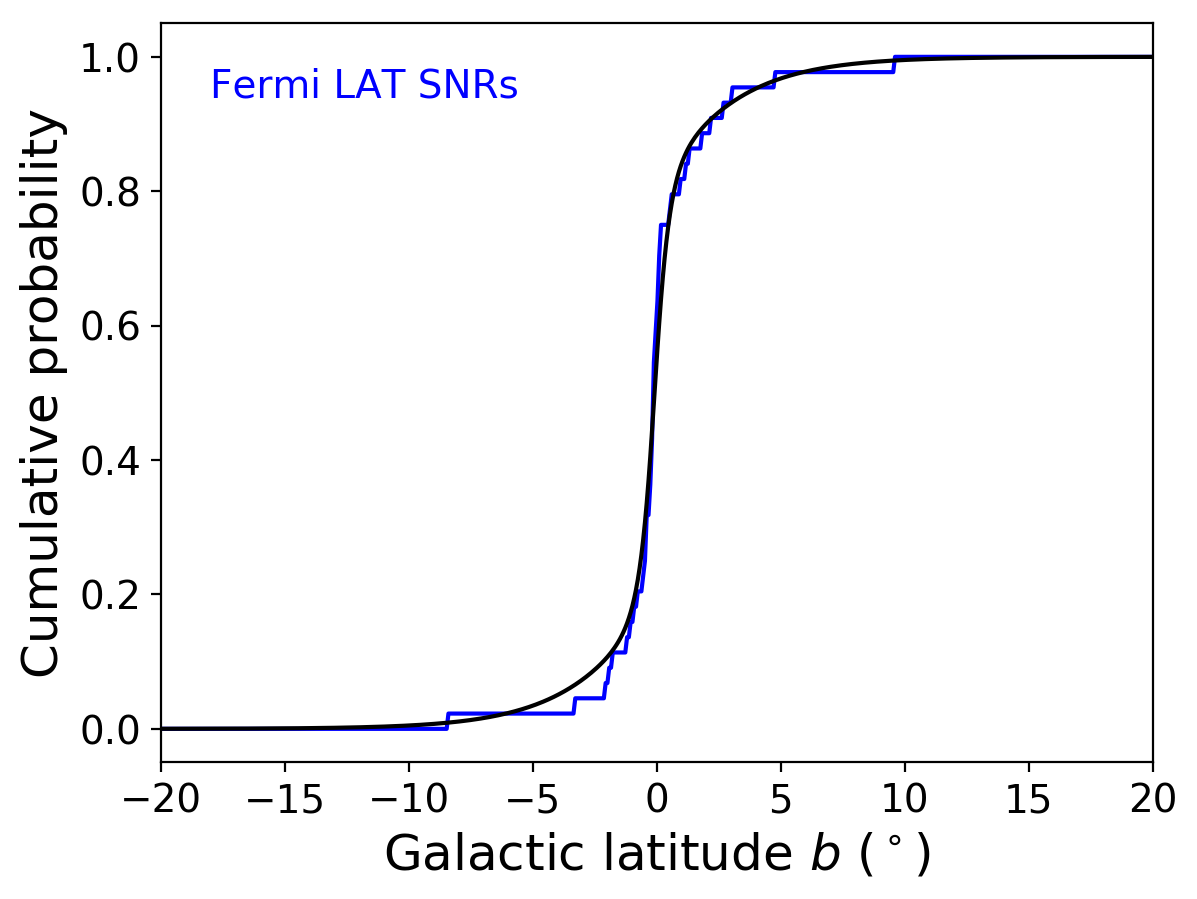}
\includegraphics[height=1.8in]{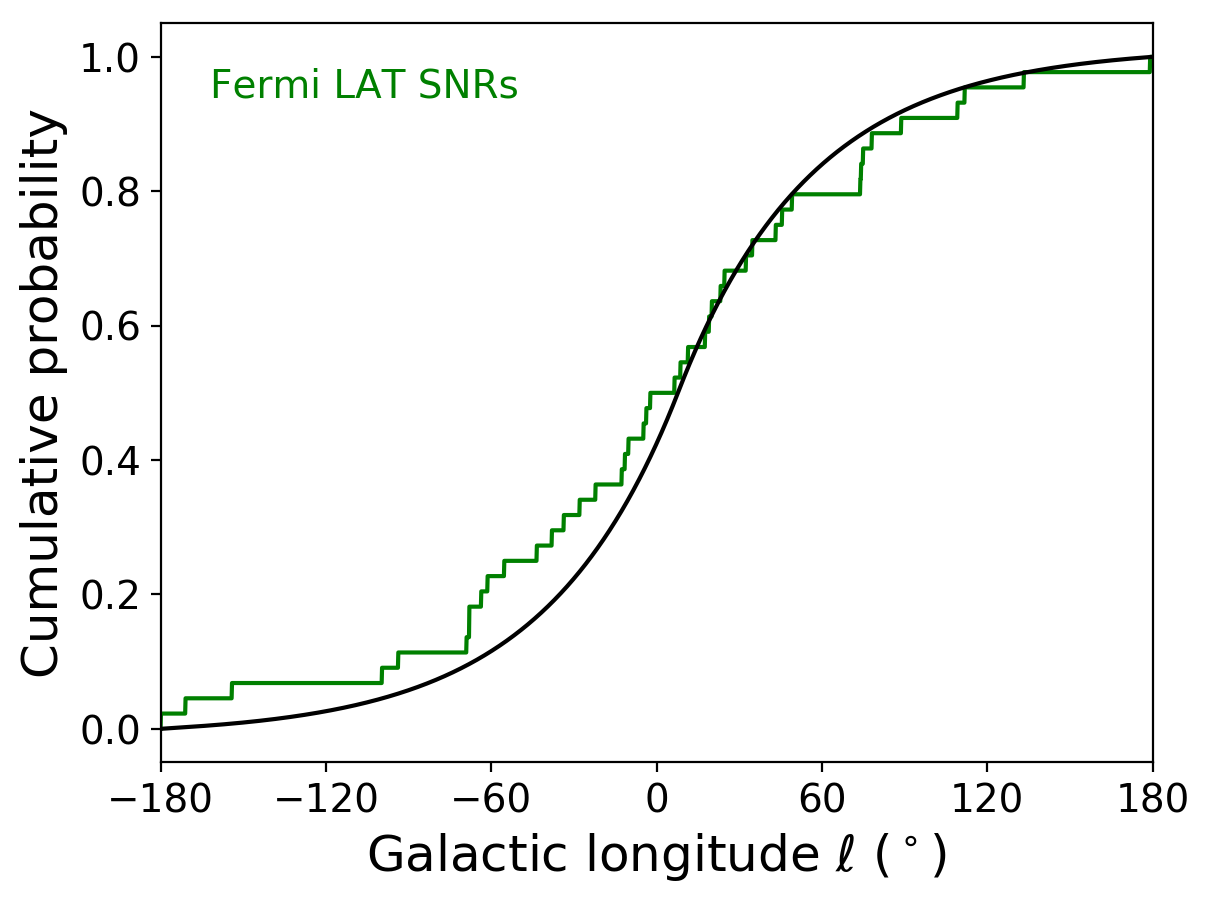}
\caption{Empirical cumulative distribution functions (ECDFs) for each catalog of candidate sources plotted against Galactic latitude (left column, blue lines) and Galactic longitude (right column, green lines).
Rows from top to bottom: OPENCLUST \cite{Dias02}, ADOC list \cite{Odrowski13}, Green SNRs \cite{Green14, Green17}, and Fermi LAT Classified and Marginally Classified SNRs \cite{Acero16}.
The black lines overlaid on each plot show the CDF of the distributions which were used to model the catalog data when generating the null distributions, as described in the text.
The horizontal axes of the Galactic latitude plots have been constrained in order to show details near the Galactic plane.}
\label{dist_plots}
\end{figure}

For each class of Galactic sources considered in this work---OCs and SNRs---we used the largest catalog of sources---i.e., OPENCLUST \cite{Dias02} and the Green SNRs \cite{Green14, Green17}, respectively---for the \verb|dynesty| distribution fitting, since the largest sample is likely to provide the best representation of the true distribution.
Both OPENCLUST and the Green SNRs involve selection effects as discussed in Section~\ref{catalogs}, but these can mostly be overlooked, since our primary goal is to produce a randomized facsimile of \textit{the catalogs themselves}, rather than a fully accurate Galactic model.
The best latitude models for both the OPENCLUST data and the Green SNRs data used the double hyperbolic secant given in Equation~\ref{lat_dist}, but the best longitude model for the OPENCLUST data was the uniform model of Equation~\ref{lon_dist_flat}, while the Green SNRs data was a better match to the exponential disk of Equation~\ref{lon_dist_exp}.
Figure~\ref{dist_plots} shows how the best distribution models visually compare with the data from all four catalogs using cumulative distribution functions (CDFs) which show the cumulative integrated area of the PDF.
The median posterior parameter values for each model as obtained by \verb|dynesty| are presented in Tables~\ref{oc_lat_table}--\ref{snr_lon_table}.

\begin{table}
\centering
\caption{Median posterior values for the four parameters in the OC latitude model (Equation~\ref{lat_dist}), including the corresponding normalization value.
For each parameter, the 68\% interval shows the 16th and 84th percentiles of posterior samples.}
\begin{tabular}{|c|l|c|c|c|}
\hline
Parameter & Description & Median Value & 68\% Interval & Units \\
\hline
$b_0$ & Center offset & $-0.349$ & $[-0.416, -0.284]$ & deg \\
$h_{b1}$ & Scale height 1 & $1.83$ & $[1.77, 1.90]$ & deg \\
$h_{b2}$ & Scale height 2 & $10.1$ & $[9.51, 10.8]$ & deg \\
$f_1$ & Amplitude fraction & $0.950$ & $[0.943, 0.956]$ & --- \\
$A_b$ & Normalization & $0.142$ & $[0.138, 0.146]$ & deg$^{-1}$ \\
\hline
\end{tabular}
\label{oc_lat_table}
\end{table}

\begin{table}
\centering
\caption{Same as Table~\ref{oc_lat_table}, but for the SNR latitude model.}
\begin{tabular}{|c|l|c|c|c|}
\hline
Parameter & Description & Median Value & 68\% Interval & Units \\
\hline
$b_0$ & Center offset & $-0.0814$ & $[-0.124, -0.0383]$ & deg \\
$h_{b1}$ & Scale height 1 & $0.384$ & $[0.335, 0.438]$ & deg \\
$h_{b2}$ & Scale height 2 & $2.58$ & $[2.28, 2.94]$ & deg \\
$f_1$ & Amplitude fraction & $0.921$ & $[0.899, 0.938]$ & --- \\
$A_b$ & Normalization & $0.568$ & $[0.517, 0.625]$ & deg$^{-1}$ \\
\hline
\end{tabular}
\label{snr_lat_table}
\end{table}

\begin{table}
\centering
\caption{Same as Tables~\ref{oc_lat_table} and~\ref{snr_lat_table}, but for the SNR longitude model (Equation~\ref{lon_dist_exp}).}
\begin{tabular}{|c|l|c|c|c|}
\hline
Parameter & Description & Median Value & 68\% Interval & Units \\
\hline
$\ell_0$ & Center offset & $7.96$ & $[5.27, 10.7]$ & deg \\
$h_\ell$ & Scale height & $48.4$ & $[45.1, 52.1]$ & deg \\
$A_\ell$ & Normalization & $0.0106$ & $[0.00991, 0.0113]$ & deg$^{-1}$ \\
\hline
\end{tabular}
\label{snr_lon_table}
\end{table}

The top two panels of Figure~\ref{dist_plots} show the best-fit models for the OPENCLUST data using Equations~\ref{lat_dist} and~\ref{lon_dist_flat}.
The largest visible discrepancies occur in the longitude distribution between $\ell \simeq -120^\circ$ and $\ell \simeq +60^\circ$.
The OCs listed in OPENCLUST are all visible at optical wavelengths, implying that they are all close enough to still be observable after accounting for interstellar dust extinction.
More than 95\% of OPENCLUST sources that have tabulated distances are within $6\ \mathrm{kpc}$ of Earth, with a thin tail extending out to ${\sim}15\ \mathrm{kpc}$.
These distances are too short to reveal large-scale asymmetries in the disk structure of the Milky Way, so we expect the longitude distribution to be uniform to first order, with a few discrepancies due to the relative positions of the local spiral arms.
Since the ADOC list also consists of Galactic OCs, we expect those 36 OCs to follow the same distribution with some additional scatter due to the small number of objects.
One complication here is that the ADOC list consists only of OCs which are north of the celestial equator, so their longitude distribution is uniform, but has a gap between $\ell \simeq -147^\circ$ and $\ell \simeq +33^\circ$ where the Galactic plane dips south of the equator.
This issue is addressed by composition of the factorized PDF $F(b, \ell)$ with an additional step distribution PDF that excludes the southern hemisphere.
The second row of Figure~\ref{dist_plots} shows an approximation of the resulting longitude CDF with this step distribution filter included---the distribution models are otherwise identical to those used for OPENCLUST.

The third row in Figure~\ref{dist_plots} shows the best models for the Green SNRs data using Equations~\ref{lat_dist} and~\ref{lon_dist_exp}.
When compared to the OC distributions, there are two important differences for this case.
Firstly, the SNR latitude distribution has a narrower spread about the Galactic plane than the OC distribution.
Comparing the $h_{b1}$ and $h_{b2}$ values in Tables~\ref{oc_lat_table} and~\ref{snr_lat_table}, the SNR distribution is narrower by a factor of ${\sim}\text{4--5}$ in both parameters.
This difference is a consequence of the scale heights of young and old stars in the Galactic disk.
Stars form in the thin disk, which has a scale height of about 280--350~pc, but older stars diffuse into the thick disk, which has a larger scale height of about 0.75--1~kpc \cite[see, e.g., Section~2.2 of][]{Sparke07}.
SNRs are the remains of high-mass stars, which live short lives and die before they have time to diffuse into the thick disk, so the SNRs are typically found near stellar formation sites in the thin disk.
The second big difference for SNRs is the form of the longitude distribution, which follows an exponential instead of a uniform distribution---there are more SNRs visible when looking towards the Galactic center than away from it \cite[for an in-depth discussion, see][]{Green15}.
This is because the Green SNR catalog was primarily compiled using observations at radio wavelengths, which detect SNRs via their synchrotron radiation.
Since long wavelengths are subject to minimal extinction from interstellar dust, large-scale asymmetries in the star formation activity of the Galactic disk are more apparent.
The fourth row of Figure~\ref{dist_plots} shows that the 44 objects from the Fermi LAT Classified \& Marginally Classified SNRs catalog also conform to the same distribution whose parameters were obtained from the Green SNRs data.
Again, the smaller number of objects produces some additional scatter.

\section{Results and Discussion} \label{results}

In this section, we present the results of our analyses.
For each candidate source set, we present a sky map in Galactic coordinates showing the arrival directions of the IceCube neutrino events (with circularized errors at the 90\% confidence level) as well as the positions of the relevant candidate sources.
We also include comparative histograms showing the empirical $r$ distribution and the Monte Carlo null distribution of $r$ as described in Section~\ref{the_method}.
The null distribution histograms display the averaged distribution over $K = 10^6$ Monte Carlo realizations.
We note that negative $r$ values can occur when the most probable arrival direction of a neutrino falls within the angular extent of a candidate source.
For our histograms, any $r < 0$ will be rounded up to $r = 0$ and then tabulated in the first bin (covering $0 \leq r < 1$), so that the first bin always represents $N_c$, the total number of coincidences.
In the interest of focusing on neutrinos with small $r$ values, the histograms are truncated to the range $r < 40$.
For each analysis, we report $\Delta N_c$, the number of excess coincidences in the empirical $r$ distribution when compared with the averaged null distribution, as well as $\pnc$, which is the $p$-value associated with $\Delta N_c$.
For each catalog, we also present the complementary results of a K-S test applied to the two histogram distributions, with its $p$-value denoted by $\pks$.
For both types of statistical analysis, we choose $\alpha = 0.01$ as our significance threshold---if $\pnc < \alpha$ or $\pks < \alpha$, then we will consider the disparity between the empirical and null distributions to possibly be indicative of a causal connection between the candidate sources and the neutrinos.

\subsection{Open Cluster Analyses} \label{oc_results}

\begin{figure}
\centering
\includegraphics[height=1.85in]{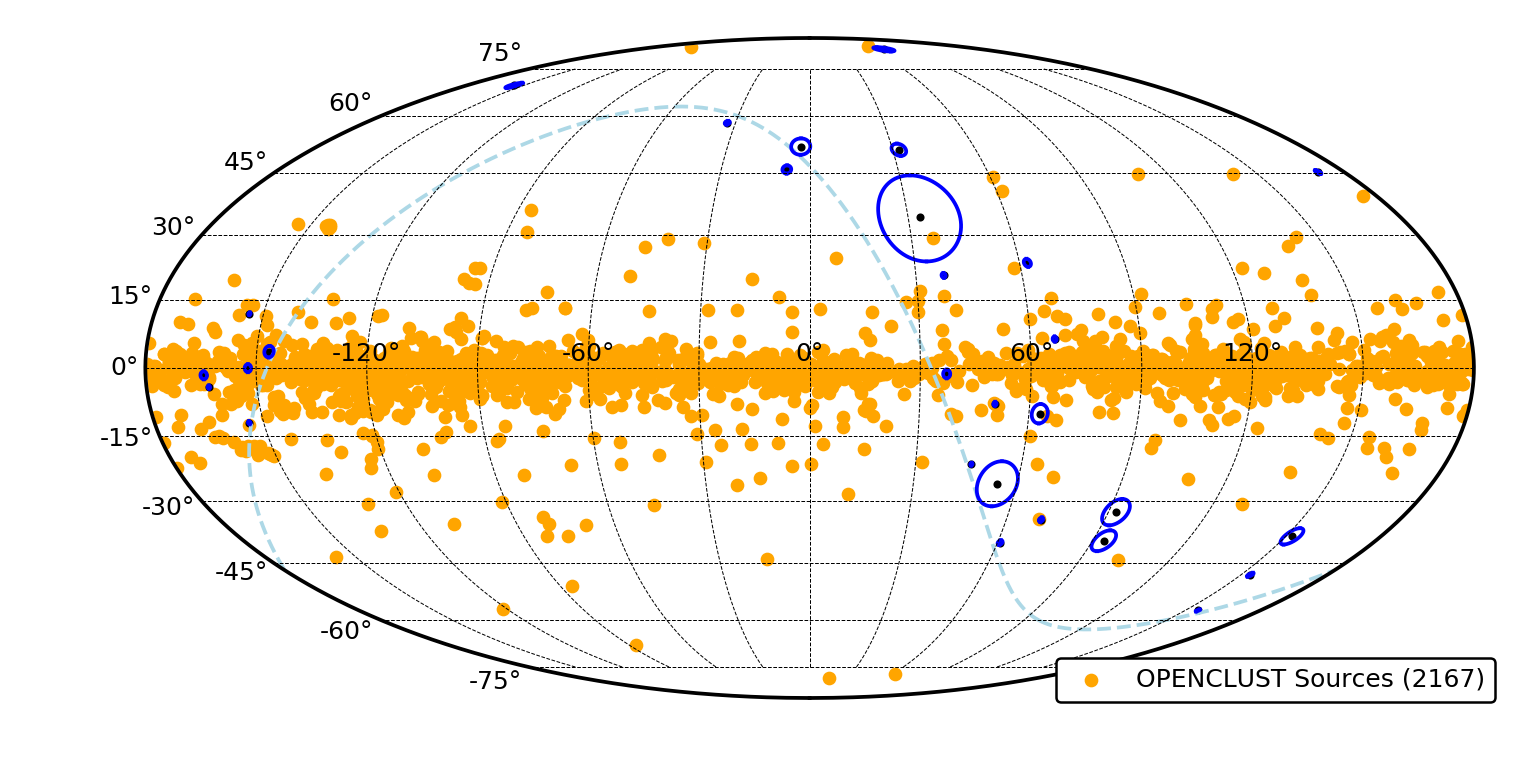}
\includegraphics[height=1.85in]{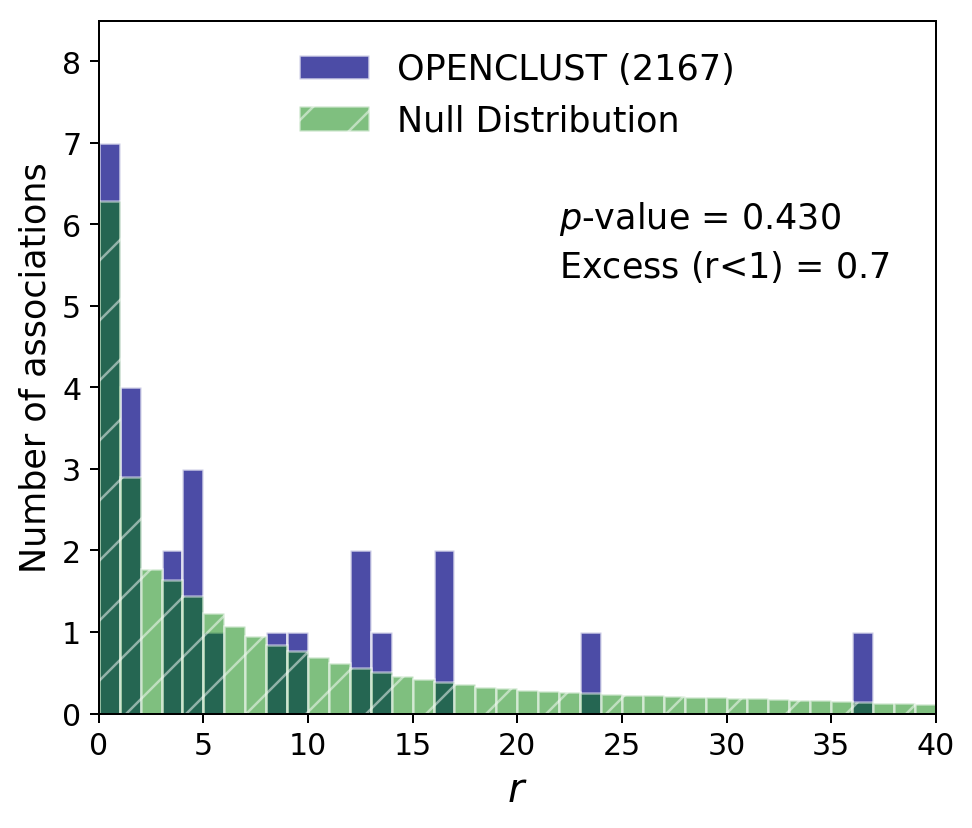}
\includegraphics[height=1.85in]{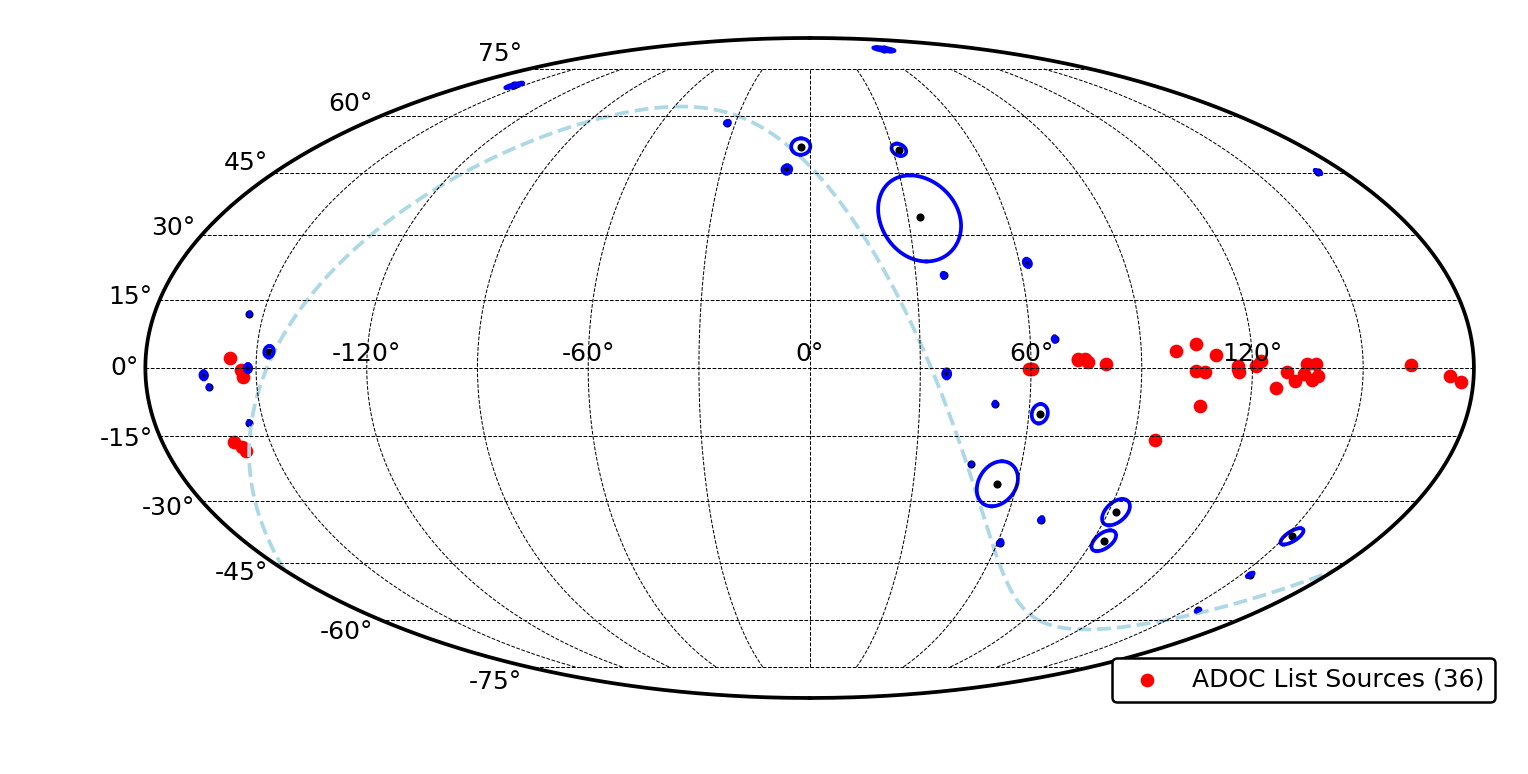}
\includegraphics[height=1.85in]{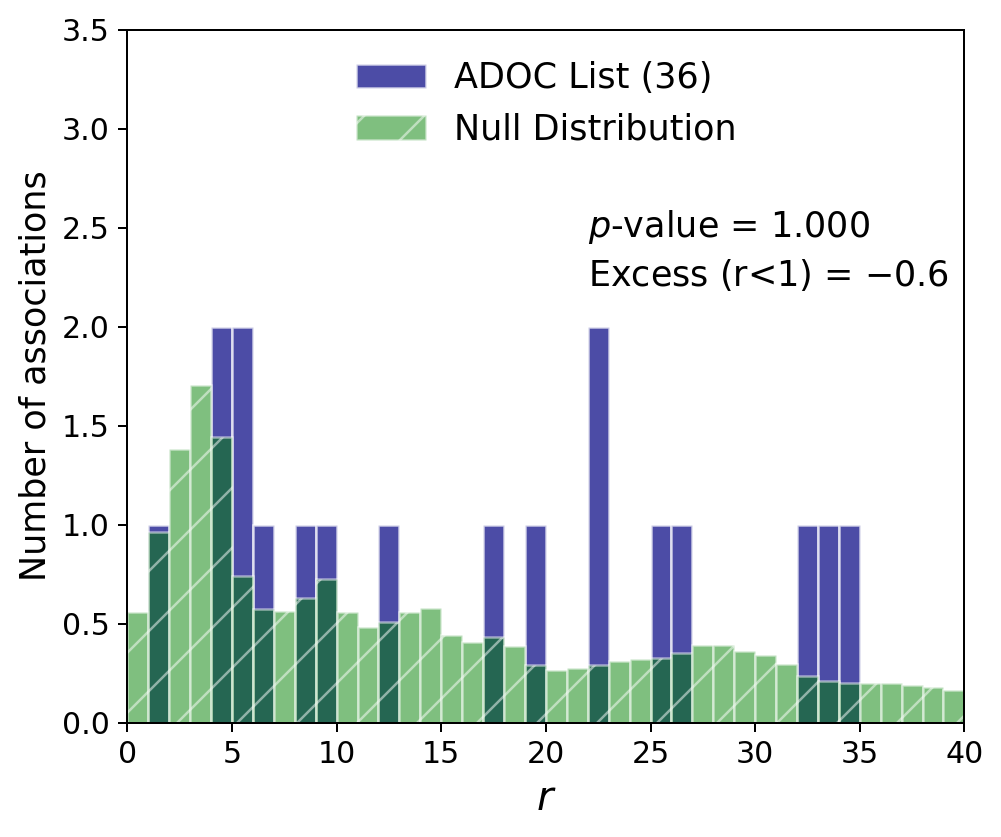}
\caption{\textit{Top left:} Sky map showing the positions of OC candidate sources (orange circles) from OPENCLUST \cite{Dias02}.
Black dots mark the arrival directions of track-like neutrinos \cite{Aartsen16}, with blue circles showing their circularized angular errors.
The dashed light blue line is the celestial equator.
\textit{Top right:} Histogram of $r$ values for the analysis of the OPENCLUST sources.
The empirical distribution of $r$ values is shown in blue, with the averaged null distribution overplotted in semi-transparent green.
See text for details.
\textit{Bottom left:} Same as top left, but showing the positions of the 36 ADOCs \cite{Odrowski13} instead (red circles).
\textit{Bottom right:} Same as top right, but showing $r$ distributions for the ADOCs analysis.}
\label{oc_figure}
\end{figure}

The top two panels of Figure~\ref{oc_figure} show the analysis results for the hypothesis of causal association with the OPENCLUST \cite{Dias02} candidate sources.
The empirical $r$ distribution has $N_c = 7$ coincidences, which is an excess of $\Delta N_c = 0.7$ coincidences over the averaged null distribution expectation.
Table~\ref{covocc_table} gives specific information regarding each of the seven coincidences, as well as the four additional cases in which $r < 3$.
While these cases are not coincidences, they are near enough to potentially be of some interest.
The fraction of Monte Carlo null distribution realizations which yielded at least seven coincidences by chance was $\pnc = 0.430$, which is far too large to be considered significant at the $\alpha = 0.01$ level.
Likewise, a K-S test comparing the two distributions resulted in $\pks = 0.973$, suggesting that the empirical $r$ distribution is in excellent agreement with the null distribution.

The two lower panels of Figure~\ref{oc_figure} show the results of our analysis using the ADOC list \cite{Odrowski13} with 36 objects.
The empirical $r$ distribution did not result in any coincidences at all, in contrast to the averaged null distribution expectation of ${\sim}0.6$ coincidences, so we find $\Delta N_c = -0.6$.
Table~\ref{adoc_table} provides a few details for the single case where $r < 3$ in this analysis, which had $r = 1.442$.
Since all null distribution realizations will necessarily produce zero or more coincidences, $\pnc = 1.0$ for this analysis.
The associated K-S test result was $\pks = 0.442$, which also greatly exceeds the significance threshold.
In summary, we find no evidence of causal association between the neutrino events and the ADOCs.

\begin{table}
\centering
\caption{List of the eleven track-like neutrino events having $r < 3$ with OPENCLUST \cite{Dias02} sources.
The dashed line separates coincidences ($r < 1$) from events with $1 < r < 3$.
The $\nu$~ID and $E_\nu$ columns list the IceCube neutrino ID number and energy proxy taken from Table~4 of \cite{Aartsen16}.
The $\sigma_\nu$ column is the circularized 90\% confidence level error on the neutrino arrival direction $b_\nu, \ell_\nu$, which is listed in Galactic coordinates.
The nearest candidate source is listed with its position $b, \ell$, angular radius $a$, and center-to-center separation $S$ from the neutrino event.}
\vspace{0.1in}
\resizebox{\textwidth}{!}{  
\begin{tabular}{|c|c|c|r|r|r|c|r|r|r|r|}
\hline
$r$ & $\nu$ ID & $E_\nu$ (TeV) & $\sigma_\nu$ ($^\circ$) & $b_\nu$ ($^\circ$) & $\ell_\nu$ ($^\circ$) & Source & $b$ ($^\circ$) & $\ell$ ($^\circ$) & $a$ ($^\circ$) & $S$ ($^\circ$) \\
\hline
0.102 & 2 & 250 & 0.438 & $-7.93$ & $+50.52$ & NGC 6837 & $-8.01$ & $+50.52$ & 0.038 & 0.083 \\
0.195 & 26 & 340 & 1.276 & $+3.61$ & $-146.67$ & NGC 2324 & $+3.30$ & $-146.55$ & 0.088 & 0.337 \\
0.364 & 28 & 210 & 0.882 & $-0.03$ & $-152.21$ & NGC 2269 & $+0.30$ & $-152.11$ & 0.025 & 0.346 \\
0.429 & 21 & 670 & 0.881 & $-1.61$ & $-164.23$ & FSR 0940 & $-1.29$ & $-164.45$ & 0.012 & 0.389 \\
0.530 & 6 & 770 & 10.248 & $+34.07$ & $+33.47$ & Dol Dzim 7 & $+29.17$ & $+36.29$ & 0.025 & 5.457 \\
0.597 & 8 & 660 & 0.530 & $-34.30$ & $+70.61$ & NGC 7193 & $-34.28$ & $+70.09$ & 0.108 & 0.425 \\
0.902 & 5 & 230 & 2.156 & $-10.13$ & $+62.97$ & NGC 6938 & $-10.74$ & $+64.91$ & 0.060 & 2.005 \\
\hdashline
1.089 & 9 & 950 & 0.340 & $-12.12$ & $-153.96$ & NGC 2112 & $-12.61$ & $-154.13$ & 0.150 & 0.520 \\
1.202 & 17 & 200 & 1.020 & $+82.97$ & $+77.71$ & Latham 1 & $+84.59$ & $+72.32$ & 0.500 & 1.726 \\
1.440 & 10 & 520 & 0.961 & $-1.31$ & $+37.15$ & NGC 6755 & $-1.69$ & $+38.60$ & 0.117 & 1.500 \\
1.996 & 14 & 210 & 5.187 & $-25.98$ & $+54.33$ & NGC 7036 & $-21.44$ & $+64.54$ & 0.033 & 10.387 \\
\hline
\end{tabular}
}  
\label{covocc_table}
\end{table}

\begin{table}
\centering
\caption{Tabulated details for the one track-like neutrino event having $r < 3$ with an ADOC \cite{Odrowski13}.
Columns are the same as in Table~\ref{covocc_table}.}
\vspace{0.1in}
\resizebox{\textwidth}{!}{  
\begin{tabular}{|c|c|c|r|r|r|c|r|r|r|r|}
\hline
$r$ & $\nu$ ID & $E_\nu$ (TeV) & $\sigma_\nu$ ($^\circ$) & $b_\nu$ ($^\circ$) & $\ell_\nu$ ($^\circ$) & Source & $b$ ($^\circ$) & $\ell$ ($^\circ$) & $a$ ($^\circ$) & $S$ ($^\circ$) \\
\hline
1.442 & 28 & 210 & 0.882 & $-0.03$ & $-152.21$ & Collinder 106 & $-0.40$ & $-153.96$ & 0.520 & 1.791 \\
\hline
\end{tabular}
}  
\label{adoc_table}
\end{table}


It is interesting to note that none of the OPENCLUST coincidences in Table~\ref{covocc_table} involved young OCs: five out of the seven coincident sources have ages of 1~Gyr or older, placing them among the oldest 25\% of OPENCLUST sources with tabulated ages.
In fact, NGC~7193 (aged 4.47~Gyr) is in the top 2\%.
The youngest two coincident OCs are NGC~2324 and NGC~2269, both of which still have ages over 200~Myr.
All eleven OCs in Table~\ref{covocc_table} are older than the threshold of ${\sim}40\ \mathrm{Myr}$ that was used to select the 36 ADOCs \cite{Odrowski13}.
NGC~6755, with $r = 1.440$ and an age of 52~Myr, is the only OC in the table that is younger than 100~Myr.

In addition to being old, the OPENCLUST sources in Table~\ref{covocc_table} are also preferentially nearby, being below the catalog's median distance of 1800~pc for OCs with tabulated distances, with the exception of NGC~2324 (3800~pc) and FSR~0940 (2421~pc).
However, the OCs in Table~\ref{covocc_table} are \textit{not} preferentially those having the largest angular sizes.
Five of the eleven OCs listed have angular diameters below the catalog median of 5.0~arcmin (radius $0.042^\circ$).
Those five OCs include FSR~0940, which is among the smallest 10\% of OCs with tabulated angular sizes due to its diameter of only 1.4~arcmin (radius $0.012^\circ$).
Despite being a comparatively small target, FSR~0940 found itself in a coincidence at $r = 0.429$ with neutrino ID 21.

OPENCLUST is such a broad catalog that many objects it includes have not been studied in detail.
Typical OCs are rarely mentioned in literature outside of large studies that survey tens to thousands of objects.
For instance, a recent study concludes that NGC~7193 and NGC~7036, both of which appear in Table~\ref{covocc_table}, are most likely to be asterisms \cite{CantatGaudin20}.
While their stars appear in the same direction on the sky, they do not form a convincing cluster in either parallax or proper motion space, so they are not likely to be physical clusters of gravitationally bound stars with a common origin.
The status of NGC~6837 is similarly dubious \cite{KroneMartins10}.
None of these eleven OCs have been explicitly considered as potential neutrino sources in past literature.

While the ADOC analysis produced no coincidences, there is one case of $r < 3$ documented in Table~\ref{adoc_table}.
The relevant ADOC, Collinder~106, is among the 10 ADOCs that are within the Local Arm of the Galaxy, and it has a larger angular size (radius $0.52^\circ$) than two-thirds of the ADOCs.
The OC is located in the Monoceros Loop within the larger Monoceros star-forming region \cite{Costado18}.
It has also been proposed that Collinder~106 is associated with the gamma-ray pulsar J0633$+$0632 \cite{Danilenko20}.
The COCD, which the ADOCs were selected from, lists Collinder~106 with an age estimate of 5.50~Myr, which would place it among the youngest 25\% of ADOCs.
However, the same source is also cataloged in OPENCLUST, where it has a dramatically different age estimate of 7.94~Gyr.
This inconsistency suggests the need for further study of this object.

There are also trends among the track-like neutrino events that appear in Tables~\ref{covocc_table}--\ref{adoc_table}.
Neutrino ID 28 appears in both tables, which may be due in part to the fact that it is the closest of the 29 neutrino events to the Galactic plane ($b_\nu = -0.03^\circ$).
This same event also has one of the lowest energy proxies in the IceCube data set---all of the track-like events were selected to have energies of at least 200~TeV, and ID~28 with $E_\nu = 210\ \mathrm{TeV}$ is close to that threshold.
Neutrino IDs 10, 21, and 26 (which all appear in Table~\ref{covocc_table}) are the next closest events to the Galactic plane, with $|b_\nu| < 4^\circ$ for their most probable arrival directions.
The closest coincidence in the OPENCLUST analysis (Table~\ref{covocc_table}) is neutrino ID~2 at $r = 0.102$ from NGC~6837.
ID~2 is within $10^\circ$ of the Galactic plane and also has a small energy proxy at $E_\nu = 250\ \mathrm{TeV}$.
The only neutrino event appearing in either Table~\ref{covocc_table} or Table~\ref{adoc_table} that has $|b_\nu| > 35^\circ$ is ID~17.
At $b_\nu = 82.97^\circ$, it is the farthest of the 29 neutrino events from the Galactic plane, making its near-coincidence with an OC something of an unusual result.
We also note that neutrino ID 6, which has an OPENCLUST coincidence (Dol~Dzim~7) in Table~\ref{covocc_table}, has the largest circularized error among the 29 neutrinos at $\sigma_\nu = 10.248^\circ$.
The correspondingly large area that its error ellipse covers on the sky increases the likelihood of its involvement in chance coincidences.

\subsection{Supernova Remnant Analyses} \label{snr_results}

\begin{figure}
\centering
\includegraphics[height=1.85in]{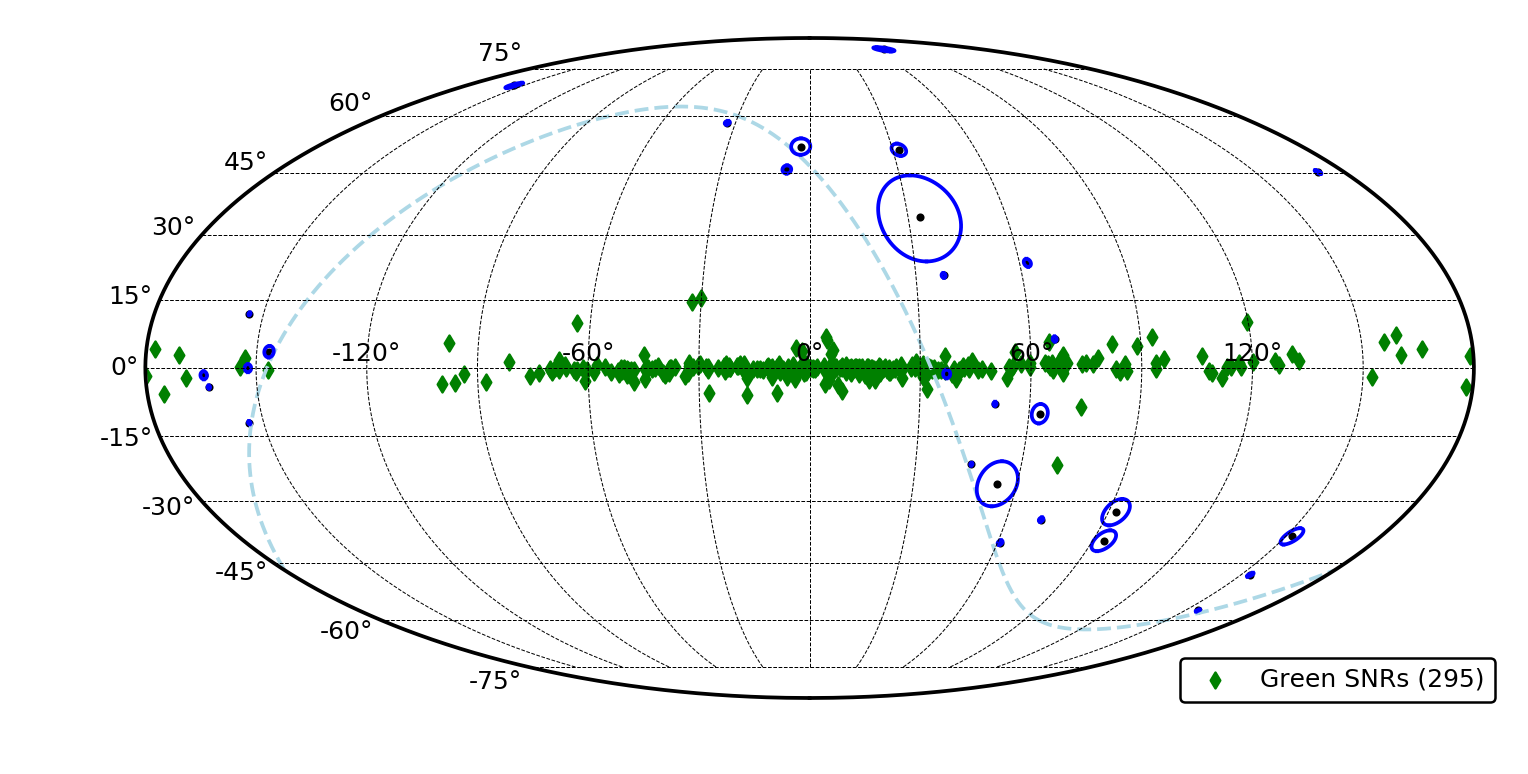}
\includegraphics[height=1.85in]{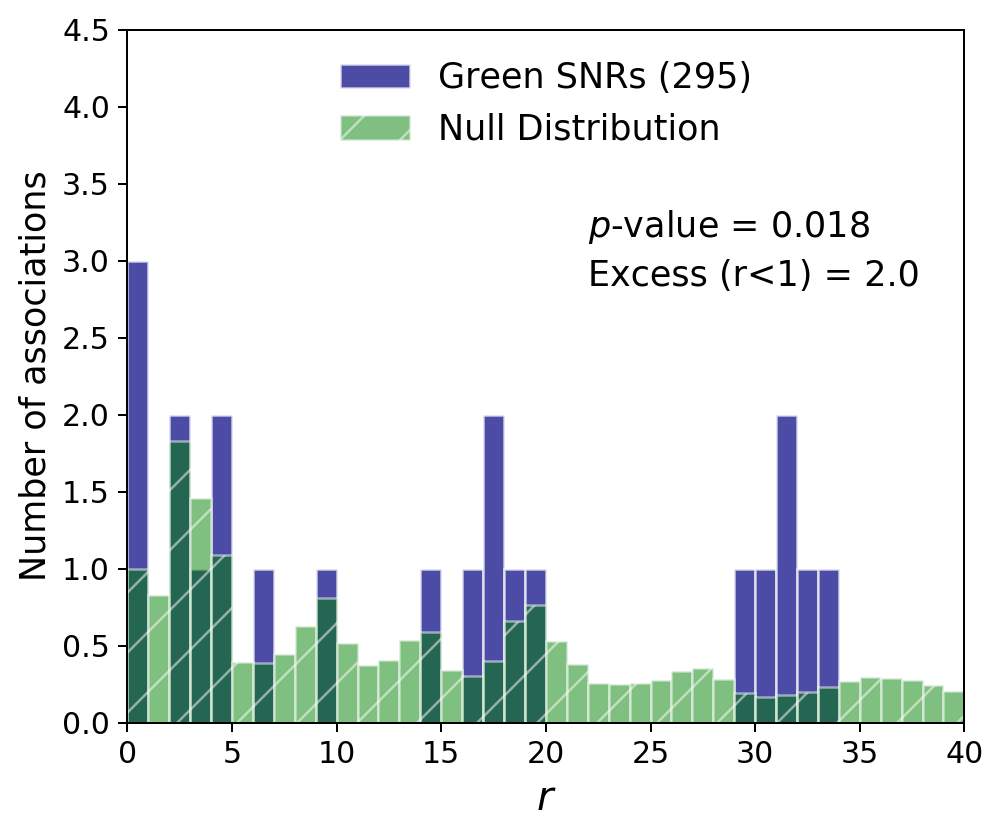}
\includegraphics[height=1.85in]{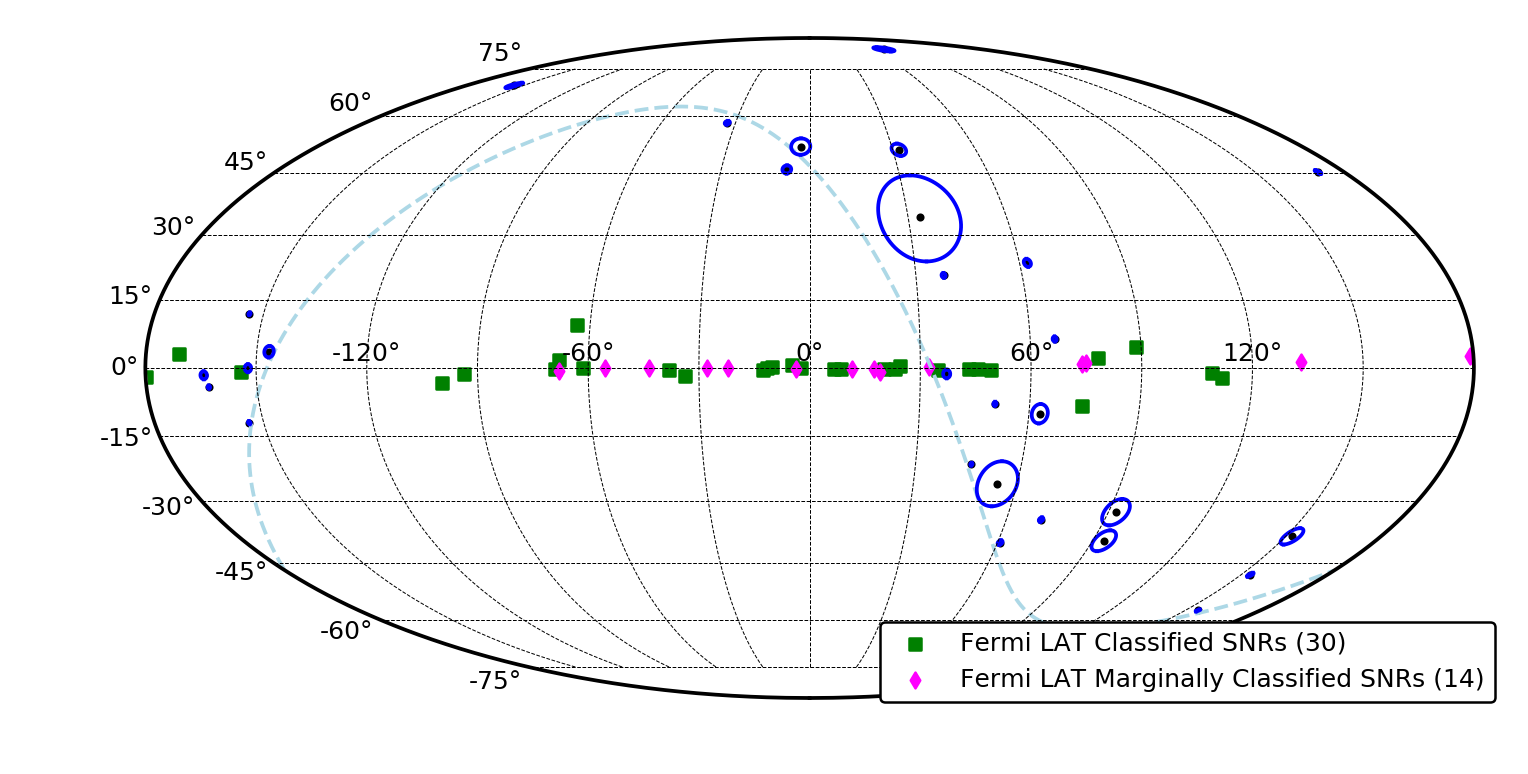}
\includegraphics[height=1.85in]{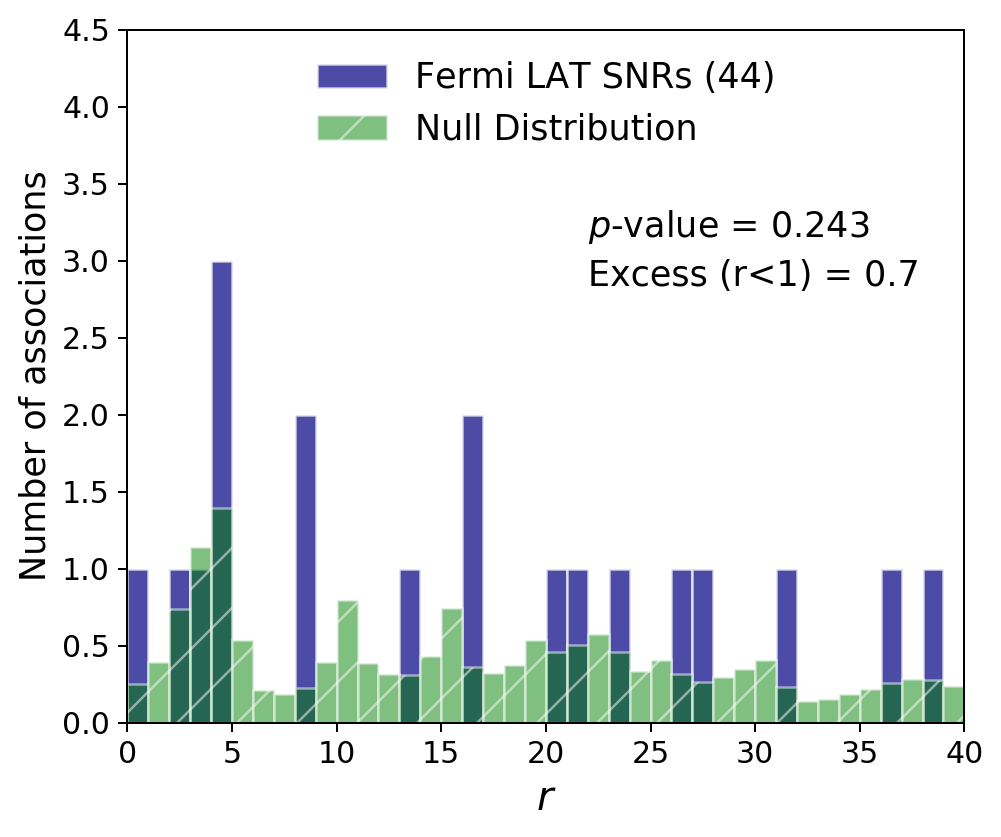}
\caption{\textit{Top left:} Same as sky maps in Figure~\ref{oc_figure}, but showing positions of the 295 Green SNRs \cite{Green14, Green17} with green diamonds.
\textit{Top right:} Same as histograms in Figure~\ref{oc_figure}, but showing the relevant $r$ distributions for the Green SNRs analysis.
\textit{Bottom left:} Sky map of the 44 Fermi LAT Classified and Marginally Classified SNRs \cite{Acero16}.
Classified SNRs are indicated with green squares, and marginally classified SNRs are indicated with magenta diamonds.
\textit{Bottom right:} Histogram of $r$ distributions for the Fermi LAT SNRs analysis.}
\label{snr_figure}
\end{figure}

The top sky map and histogram in Figure~\ref{snr_figure} show the analysis results when using the Green SNRs catalog \cite{Green14, Green17} as the list of candidate sources.
There were $N_c = 3$ coincidences in the empirical $r$ distribution, yielding an excess of $\Delta N_c = 2.0$ over the null expectation of only about 1.0 coincidence.
The three coincidences are listed with details in Table~\ref{green_table} along with two additional cases where $r < 3$.
The $p$-value was $\pnc = 0.018$, which is close to being significant at the threshold level of $\alpha = 0.01$.
Alternatively, the result of a K-S test comparing the two distributions was $\pks = 0.444$, which leads us to conclude that this case is also consistent with a null distribution.

The lower sky map and histogram in Figure~\ref{snr_figure} show the analysis results for the Fermi LAT Classified and Marginally Classified SNRs \cite{Acero16}.
The number of coincidences present was $N_c = 1$, which is an excess of $\Delta N_c = 0.7$ over the null distribution expectation.
The one coincidence is detailed in Table~\ref{fermar_table}, as is one other case of $r < 3$ occurring in this analysis.
The fraction of null distribution iterations producing at least one coincidence was $\pnc = 0.243$, which is well above the threshold.
The K-S test result of $\pks = 0.699$ for the two distributions similarly indicates consistency between the data and the Monte Carlo null case.

\begin{table}
\centering
\caption{List of the five track-like neutrino events having $r < 3$ with Green SNRs \cite{Green14, Green17}.
The dashed line separates coincidences ($r < 1$) from events with $1 < r < 3$.
Columns are the same as in Table~\ref{covocc_table}.}
\vspace{0.1in}
\resizebox{\textwidth}{!}{  
\begin{tabular}{|c|c|c|r|r|r|c|r|r|r|r|}
\hline
$r$ & $\nu$ ID & $E_\nu$ (TeV) & $\sigma_\nu$ ($^\circ$) & $b_\nu$ ($^\circ$) & $\ell_\nu$ ($^\circ$) & Source & $b$ ($^\circ$) & $\ell$ ($^\circ$) & $a$ ($^\circ$) & $S$ ($^\circ$) \\
\hline
$-1.116$ & 24 & 850 & 0.555 & $+6.42$ & $+66.66$ & G65.3$+$5.7 & $+5.66$ & $+65.18$ & 2.273 & 1.654 \\
$\phantom{-}0.270$ & 28 & 210 & 0.882 & $-0.03$ & $-152.21$ & G205.5$+$0.5 & $+0.21$ & $-154.27$ & 1.833 & 2.072 \\
$\phantom{-}0.649$ & 10 & 520 & 0.961 & $-1.31$ & $+37.15$ & G36.6$-$0.7 & $-0.69$ & $+36.59$ & 0.208 & 0.832 \\
\hdashline
$\phantom{-}2.140$ & 26 & 340 & 1.276 & $+3.61$ & $-146.67$ & G213.0$-$0.6 & $-0.36$ & $-146.69$ & 1.247 & 3.977 \\
$\phantom{-}2.444$ & 14 & 210 & 5.187 & $-25.98$ & $+54.33$ & G70.0$-$21.5 & $-21.54$ & $+70.03$ & 2.345 & 15.023 \\
\hline
\end{tabular}
}  
\label{green_table}
\end{table}

\begin{table}
\centering
\caption{List of the two track-like neutrino events having $r < 3$ with Fermi LAT SNRs \cite{Acero16}.
The dashed line separates the coincidence ($r < 1$) from the event with $1 < r < 3$.
Columns are the same as in Table~\ref{covocc_table}.
Both candidate sources appearing in this table are listed among the 30 classified SNRs in \cite{Acero16} (as opposed to the 14 marginally classified SNRs).}
\vspace{0.1in}
\resizebox{\textwidth}{!}{  
\begin{tabular}{|c|c|c|r|r|r|c|r|r|r|r|}
\hline
$r$ & $\nu$ ID & $E_\nu$ (TeV) & $\sigma_\nu$ ($^\circ$) & $b_\nu$ ($^\circ$) & $\ell_\nu$ ($^\circ$) & Source & $b$ ($^\circ$) & $\ell$ ($^\circ$) & $a$ ($^\circ$) & $S$ ($^\circ$) \\
\hline
$-0.276$ & 28 & 210 & 0.882 & $-0.03$ & $-152.21$ & G205.5$+$0.5 & $-0.82$ & $-154.09$ & 2.280 & 2.037 \\
\hdashline
$\phantom{-}2.410$ & 10 & 520 & 0.961 & $-1.31$ & $+37.15$ & G34.7$-$0.4 & $-0.44$ & $+34.66$ & 0.310 & 2.627 \\
\hline
\end{tabular}
}  
\label{fermar_table}
\end{table}


A noteworthy detail in regards to these SNR results is that the associated candidate sources are preferentially those that cover larger areas on the sky, which is suggestive of chance coincidences.
The median circularized angular diameter of the entire catalog of Green SNRs is 17~arcmin (radius $0.14^\circ$).
Of the six total SNRs appearing in Tables~\ref{green_table}--\ref{fermar_table} (G205.5$+$0.5 appears twice), all have circularized angular sizes larger than the median, and four of the six are among the largest 5\% of objects in the Green SNR catalog.
In particular, G70.0$-$21.5 is actually the largest Green SNR in the catalog, having a circularized angular diameter of 280~arcmin (radius $2.3^\circ$).
Interestingly, the same SNR also has the distinction of being located at $|b| = 21.5^\circ$ from the Galactic plane, which is the most extreme Galactic latitude of any Green SNR.

One more notable result is the appearance of G34.7$-$0.4 in the Fermi LAT SNRs analysis (Table~\ref{fermar_table}).
Better known as W44, this source is one of two SNRs whose gamma-ray spectra were found to display a characteristic pion-decay signature, providing evidence of cosmic ray proton acceleration in SNRs \cite{Ackermann13}.
This SNR also appears to be interacting with nearby molecular clouds \cite{Claussen97, Chevalier99}.
At $r = 2.410$, our analysis does not show a coincidence involving this object, but W44 is not the only SNR in Tables~\ref{green_table}--\ref{fermar_table} to have been previously considered as a cosmic ray accelerator.
The coincident SNR G65.3$+$5.7 has been considered as a potential cosmic ray accelerator in the past \cite[e.g.,][]{Kobayashi04, Delahaye10}, but it is suspected to be a \textit{leptonic} cosmic ray source, which would be inconsistent with the pion-decay model for production of high-energy neutrinos.
G205.5$+$0.5 has also been studied as a particle accelerator candidate \cite[e.g.,][]{Fiasson08}, as well as for possible interactions with the nearby Rosette Nebula \cite[e.g.,][]{Delahaye10, Xiao12, Su17}.
G213.0$-$0.6 is also a candidate for interaction with molecular clouds in its proximity \cite[e.g.,][]{Su17}.

Both of the SNR analyses found a coincidence between neutrino ID 28 and the candidate source G205.5$+$0.5, also known as the Monoceros Nebula, Monoceros Loop, or Monoceros SNR.
Of the five SNRs appearing in Table~\ref{green_table} for the Green SNRs analysis, G205.5$+$0.5 is the only one to appear in the Fermi LAT SNRs catalog \cite{Acero16} as either a classified or marginally classified SNR.
The gamma-ray source detected by Fermi LAT above 1~GeV has an apparent angular radius of ${\sim}2.3^\circ$, nearly 30\% larger than the ${\sim}1.8^\circ$ radius observed in the radio \cite{Acero16}, which accounts for most of the difference in $r$ values observed between our two analyses.
The gamma-ray source is also offset slightly in the direction of the Rosette Nebula \cite{Acero16}, which slightly reduces the separation angle $S$ between it and the neutrino arrival direction.
Since the publication of \cite{Acero16}, there has also been a detailed analysis of the GeV morphology of this SNR by Fermi LAT which concluded that the decay of neutral pions could explain the observed gamma-ray emission \cite{Katagiri16}.

As with the OC analyses, we close this discussion by examining trends in the track-like neutrino events that appear in Tables~\ref{green_table}--\ref{fermar_table}.
Neutrino ID~28 is notable for having appeared in all four analyses, coincident with G205.5$+$0.5 in both SNR analyses, coincident with NGC~2269 in the OPENCLUST analysis, and having $r = 1.442$ with Collinder~106 in the ADOC analysis.
Also notable is neutrino ID~10, which appears in three out of four analyses, though only in the Green SNRs analysis (Table~\ref{green_table}) was it close enough to a candidate source to have a coincidence (SNR G36.6$-$0.7).
In the Fermi LAT SNRs analysis (Table~\ref{fermar_table}), ID~10 was associated with a different SNR and only at $r = 2.410$.
Neutrino IDs~26 and 14 also appeared in more than one analysis---both appeared in the OPENCLUST analysis (Table~\ref{covocc_table}), though only ID~26 had a coincidence (NGC~2324), and both appeared without coincidences in the Green SNRs analysis (Table~\ref{green_table}).
We note again that neutrino IDs~28, 10, and 26 all have arrival directions within $4^\circ$ of the Galactic plane, which makes chance coincidences with Galactic candidate sources more likely.
In a similar vein, neutrino ID~14 has the second-largest circularized neutrino error at $\sigma_\nu = 5.187^\circ$.
The closest coincidence in the Green SNRs analysis (Table~\ref{green_table}) was neutrino ID~24, which has $r = -1.116$ with SNR G65.3$+$5.7 and appears in no other analyses.
ID~24 is still within $10^\circ$ of the Galactic plane, and is unusual in that it has one of the highest energy proxies in the track-like neutrino data set at $E_\nu = 850\ \mathrm{TeV}$.

\section{Conclusions} \label{conclusions}

Neutrino astronomy is a growing field which is expected to help illuminate the origins of cosmic rays above the knee and provide insights into the physical mechanisms powering their accelerators.
As the field matures, diverse approaches are being developed to address questions regarding the origins of the high-energy neutrinos observed by IceCube.
One fundamental question to be answered is that of a possible Galactic component, i.e., whether there exists some non-negligible fraction ($\gtrsim 5\%$) of the neutrino flux which can be attributed to the Milky Way.

In our previous work \cite{Emig15, Lunardini19}, we adopted the ``nearest neighbor'' method to search for positional associations (or \textit{coincidences}) of IceCube neutrino events with candidates from specific classes of extragalactic astronomical objects.
In the present paper, we developed this method further in order to accommodate candidate neutrino sources that are not uniformly distributed across the sky.
This new version of the method was applied to search for IceCube track-like neutrino coincidences with specific classes of Galactic objects, namely open clusters (OCs) and supernova remnants (SNRs).
As in the past, we assess the compatibility of our results with the null hypothesis of no causal relationship using Monte Carlo randomization and Kolmogorov-Smirnov tests.

In particular, this work presents the first statistical analysis of Galactic OCs as candidate neutrino sources.
Using a catalog of 2,167 OCs observed at optical wavelengths \cite{Dias02}, seven coincidences were found, a result that is statistically consistent with the null case.
The overall distribution of the $r$ parameter, which represents the normalized angular distance from a neutrino to the nearest candidate source, is consistent with the null hypothesis as well.
The analysis was then repeated with a restricted set of 36 Galactic ``accelerator-dominated'' open clusters (ADOCs) \cite{Odrowski13} which were identified as likely neutrino producers on theoretical grounds.
Zero coincidences were found and the overall $r$ distribution was again consistent with the null hypothesis.

We have identified the coincident neutrino events and OCs appearing in both of these analyses (Tables~\ref{covocc_table}--\ref{adoc_table}).
It is interesting to note that our results defy the physical motivations that were used to select the 36 ADOCs---none of the ADOCs were coincident with a neutrino event, and the coincident objects from the larger OC catalog were far older than the ADOC age threshold of ${\sim}40\ \mathrm{Myr}$.
This fact can perhaps be regarded as secondary confirmation that the coincidences are consistent with random chance.
While at this time there is no evidence that these OCs are Galactic high-energy neutrino sources, they might be worth further monitoring in the future.

In our analysis of Galactic SNRs, we found three coincidences between neutrinos and the 295 objects in Green's catalog \cite{Green14, Green17}, the most complete catalog of Galactic SNRs available.
This result was only narrowly consistent with the null case, but the overall distribution of the $r$ parameter was in good agreement the null hypothesis expectation.
As with the Galactic OC analysis, we also took the opportunity to investigate a smaller set of SNRs which were selected to be more likely neutrino sources.
When restricting our analysis to 44 SNRs observed in gamma rays by Fermi LAT \cite{Acero16}, only one coincidence was found, and again the final results were consistent with the null hypothesis.

In Tables~\ref{green_table}--\ref{fermar_table} we have identified those neutrino events and SNRs which produced coincidences in each SNR analysis.
Once again, we find that the results are in tension with the scenario of hadronic neutrino production, in which we expect gamma-ray counterparts due to $\pi^0$ decay.
G205.5$+$0.5 was the only SNR in the Fermi LAT catalog found to produce a coincidence in our analyses.
The other coincident SNRs from Green's catalog have not been detected at gamma-ray energies.
Our SNR results are overall consistent with the null results from other works, in particular the stacking analysis performed by the IceCube collaboration using 7 years of data \cite{Aartsen17b}, where $p$-values of 0.25 and higher were obtained for three different sets of 4--10 SNRs each.
Those SNRs were selected based on age and TeV gamma ray observation, and were sorted into sets based on environment (e.g., molecular cloud, pulsar wind nebula).

More broadly, our Galactic neutrino findings are compatible with the general conclusion of the IceCube Collaboration that less than 14\% of the astrophysical flux is due to Galactic sources\cite{Aartsen17b}.
Across our four analyses, we found a total of 11 coincidences involving 9 unique neutrino events from the data set of 29 track-like neutrinos \cite{Aartsen16}.
If we assume that all these neutrinos are of Galactic origin, then we can estimate the Galactic component due to OCs and SNRs at $9 / 29 \sim 31\%$.
This assumption is too generous to be realistic, but it does provide a qualitative upper limit.

Future data from IceCube will likely yield improved limits on the Galactic component.
More direct measurements of the exact fraction may also be forthcoming once sufficient statistical power is available.
This would also allow greater scrutiny into the question of which types of Galactic sources are most responsible for the neutrino flux.
However, if the Galactic component is found to be negligible, this would also have important implications for Galactic OCs and SNRs, which are theoretically well-motivated as particle accelerators.
The lack of any observable Galactic neutrino flux would certainly warrant a theoretical investigation.


\acknowledgments

GSV acknowledges support from NSF award 1615575.
CL acknowledges support from NSF awards PHY-1613708 and PHY-2012195.
RAW acknowledges support from NASA JWST Interdisciplinary Scientist grants NAG5-12460, NNX14AN10G, and 80NSSC18K0200 from GSFC.

\bibliographystyle{JHEP}
\bibliography{galactic_paper}

\end{document}